\documentclass[prb,preprint]{revtex4-1} 

\usepackage{amsmath}  
\usepackage{amsfonts} 
\usepackage{graphicx}
\usepackage{xcolor}
\usepackage{csquotes}
\usepackage{empheq}
\usepackage{fancyhdr}
\setcitestyle{authoryear,open={(},close={)}}
\usepackage[colorlinks=true,linkcolor=blue,citecolor=blue]{hyperref}%

\usepackage{nomencl}		
\makenomenclature

\graphicspath{{figures/}}

\newcommand{\kk}{k}						
\newcommand{\tp}{T}						
\newcommand{\gm}{\gamma}				
\newcommand{\hm}{H}						
\newcommand{\cc}{J}						
\newcommand{\adj}{A_{ij}}				
\newcommand{\hh}{h}						
\newcommand{\si}{s_i}					
\newcommand{\sj}{s_j}					
\newcommand{\nn}{N}						
\newcommand{\mm}{m}						
\newcommand{\kb}{k_B}					
\newcommand{\mg}{M}						
\newcommand{\pij}{p_{ij}}				
\newcommand{\gs}{[0,1]}					
\newcommand{\ms}{[-1,1]}				
\newcommand{\hmf}{H_{MF}}				
\newcommand{\heff}{h^{\mathrm{eff}}}	
\newcommand{\z}{Z}						
\newcommand{\bc}{h_c}					
\newcommand{\ns}{s_{i}^{\prime}}		
\newcommand{\nh}{h^{\prime}}		
\newcommand{\cm}{J_{ij}}				
\newcommand{\bt}{\beta}					
\newcommand{\nsj}{s_{j}^{\prime}}		
\newcommand{\mk}{\bar{k}}						
\newcommand{\tc}{T_c}					
\newcommand*{\fig }{Fig.} 
\newcommand*{\figs }{Figures} 
\newcommand*{\sect }{Sec.} 
\newcommand*{\eq }{Eq.} 
\newcommand*{\app }{Appendix}
\newcommand*{\subsect }{Subsec.} 
\newcommand*{\eqs }{Eqs.} 
\newcommand*{\ie}{i.e.}
\newcommand*{\cf}{cf.}

\newcommand*{\banet}{Barab\'{a}si-Albert Network}
\newcommand*{\bamod}{Barab\'{a}si-Albert Model}

\newcommand*{\mim}{modified Ising model}

\begin{document}
\title{A Modified Ising Model of Barab\'{a}si-Albert Network with Gene-type Spins}

\author{Jeyashree Krishnan}
\email{Correspondence: krishnan@aices.rwth-aachen.de} 
\altaffiliation[permanent address: ]{MTZ, Pauwelstrasse 19, Level 3, D-52074, Aachen, Germany} 
\affiliation{Aachen Institute for advanced study in Computational Engineering Science(AICES) Graduate School, RWTH Aachen University, Germany}
\affiliation{Joint Research Center for Computational Biomedicine(JRC-Combine), RWTH Aachen University, Germany}

\author{Reza Torabi}
\email{rezatorabi@aut.ac.ir}
\affiliation{Department of Physics and Astronomy, University of Calgary, Calgary, Alberta, Canada}

\author{Edoardo Di Napoli}
\email{dinapoli@aices.rwth-aachen.de}
\affiliation{Aachen Institute for advanced study in Computational Engineering Science(AICES) Graduate School, RWTH Aachen University, Germany}
\affiliation{J\"{u}lich Supercomputing Center, Forschungszentrum J\"{u}lich, J\"{u}lich, Germany}

\author{Andreas Schuppert}
\email{schuppert@aices.rwth-aachen.de}
\affiliation{Aachen Institute for advanced study in Computational Engineering Science(AICES) Graduate School, RWTH Aachen University, Germany}
\affiliation{Joint Research Center for Computational Biomedicine(JRC-Combine), RWTH Aachen University, Germany}

\maketitle 
\clearpage
\section*{Abstract}

The central question of systems biology is to understand how individual components of a biological system cooperate in emerging phenotypes resulting in the evolution of diseases. The disease-related phenotypes are driven by the mutual interactions of thousands of molecular entities of a similar type, such as genes or proteins. As living cells are open systems in quasi-steady state type equilibrium in continuous exchange with their environment, it has been hypothesized that computational techniques that have been successfully applied in statistical thermodynamics to describe phase transitions may provide new insights to emerging behavior of biological systems. 

In contrast to interaction networks in physics, the topology of biological interaction networks is characterized by almost scale-free network topologies with around $\approx 10^4$ to $10^5$  nodes without apparent invariance groups. Hence, system-size related effects may affect the phase transitions on the one side, but methods using invariance groups as well as brute force calculations to calculate the sum over the states is rarely applicable to biological systems. Here we will systematically evaluate the translation of computational techniques from solid-state physics to network models that closely resemble biological interaction networks and develop specific translational rules to tackle the finite size problem, the topology problem and the challenge of the necessary reduction of complexity for the scale-free network topologies.

Owing to the high degree of uncertainty of the detailed biological mechanisms driving the respective networks in cells, we will focus our analysis on logic models exhibiting only two states in each network node. Motivated by the apparent asymmetry between biological states where an entity, such as a gene exhibits boolean states \ie\ is active or inactive, we present here an adaptation of symmetric Ising model towards an asymmetric one fitting to living systems herein referred to as the \mim\ with gene-type spins. We analyze phase transitions in asymmetric Ising models by Monte Carlo simulations and propose mean-field solution of \mim\ of a network type that closely resembles real-world network, the \bamod\ of scale-free networks. We show that asymmetric Ising models show similarities to symmetric Ising models with the external field and undergoes a discontinuous phase transition of the first order and exhibits hysteresis.

Further, we show that the \mim\ can be mapped to the classical Ising model of a \banet. The simulation setup presented herein can be directly used for any biological network connectivity dataset and is also applicable for other networks that exhibit similar states of activity. This is a general statistical method to deal with non-linear large scale models arising in the context of biological systems and is scalable to any network size.

\vspace{1cm}

\textit{Keywords:} Phase transitions, Ising model, Complex networks, \banet, MCMC, Mean-Field approximations

\clearpage
\mbox{}
 
\nomenclature{$\kk$}{Node Degree}
\nomenclature{$\tp$}{Temperature}
\nomenclature{$\gm$}{Scale-free exponent}
\nomenclature{$\hm$}{Hamiltonian}
\nomenclature{$\cc$}{Coupling constant}
\nomenclature{$\adj$}{Adjacency matrix}
\nomenclature{$\hh$}{Magnetic field}
\nomenclature{$\nn$}{Network size}
\nomenclature{$\mm$}{Number of preferentially attached links}
\nomenclature{$\kb$}{Boltzmann constant}
\nomenclature{$\mg$}{Order parameter}
\nomenclature{$\pij$}{Probability that new node is linked to existing node}
\nomenclature{$\gs$}{Modified Ising spins}
\nomenclature{$\hmf$}{Mean-field hamiltonian}
\nomenclature{$\heff$}{Effective magnetic field}
\nomenclature{$\z$}{Partition function}
\nomenclature{$\ms$}{Classical spins}
\nomenclature{$\bc$}{Critical magnetic field}
\nomenclature{$\bt$}{Inverse of temperature}
\nomenclature{$\mk$}{Mean degree}
 
\printnomenclature
\clearpage
\section{Introduction} 
\label{sec:intro}

The Ising model is one of the simplest and most frequently studied models of cooperative phenomena in statistical mechanics \citep{ising_1925}. The classical Ising model is a pairwise interacting two-state system proposed to explain the structure and properties of ferromagnetic materials and has been solved exactly for one- and two-dimensional lattices \citep{Onsager1944}. In the Ising model of a two-dimensional lattice, each site carries a spin which may be up or down, and neighboring spins prefer to be parallel to each other. The external field prefers to orient the spins in the direction of the field. The spins align in the same direction at low temperature, and the system exhibits spontaneous magnetization. At high temperatures, the spins align randomly, and the system is paramagnetic.

Since then, Ising models have been extended to study phase transitions occurring in more complicated topologies such as random, small-world and scale-free networks \citep{dorogovstev_2002, pekalski_2001, barrat_2000, ferreira_2010, herrero_2008,herrero_2002, gitterman_2000, lopes_2004, albert_2002,  bianconi_2002}. For example, Ising models of networks can explain how the opinion of the individual is influenced by their contacts wherein spin up/ spin down correspond to two opposing opinions of people on a given subject \citep{aleksiejuk_2002, bartolozzi_2006, castellano_2009, contucci_2007}. Further real-world applications of Ising models of networks include socioeconomic problems such as racial segregation in the US, group herding, human culture, and language dynamics, nettle’s language change \citep{stauffer_2006}; phase transitions in neural networks \citep{aldana_2004}; communication in the World Wide Web \citep{kumar_2000}; and systems biology\citep{may_2001, pastor_2015, pastor_2001}. Hence the reductionist approach in statistical physics has led to its applications in diverse interdisciplinary fields. 

In this regard, the analogy between phase transitions occurring in living systems (such as normal to diseased state transition) and physical systems (such as condensation of water) has been well-motivated \citep{davies_2011, holstein_2013, trefois_2015, smith_2010}. The normal state to cancer state transition has been described as a process similar to the first-order irreversible discontinuous phase transition occurring in physical systems \citep{facciotti_2013, jin_2017, liu_2013, mojtahedi_2016, torquato_2010}. The central idea is that living systems are open systems in quasi-steady state type equilibrium in continuous exchange with their environment wherein cells behave like a network in heat bath under external perturbations \citep{pastor_2015, scheffer_2012}. They survive by exporting entropy to the environment in exchange for structural order, and when a control parameter increases entropy, it causes collective flipping of states which drives the system to an unstable critical state (or diseased state) thereby leading to phase transitions in living systems. In an Ising model, such a control parameter could be temperature or magnetic field, which, after a certain critical value, may cause the system to undergo a phase transition.

Hence it has been hypothesized that the translation of computational techniques that have been successfully applied in statistical thermodynamics to describe the evolution of emerging patterns as phase transitions in non-living systems may provide new insights to emerging behavior of biological systems. However since in contrast to complex interaction networks in physics the topology of biological interaction networks is characterized by almost scale-free network topologies, the computational techniques in solid-state physics requiring invariance groups in the interaction network topology e.g., translational invariance, periodicities or symmetries are not directly applicable to biological systems.  However, the size of the biological networks (usually of the order of $10^4$ to $10^5$) is very small compared to structures in solid-state physics. On the one hand, such size-related efforts may not be neglected, but on the other hand, it is far too large for a brute force calculation of the sum over the states as well.

Here we will systematically evaluate the translation of computational techniques from solid-state physics to develop specific translational rules to tackle the finite size problem, the topology problem and the challenge of the necessary reduction of complexity for the scale-free network topologies. Because of the high degree of uncertainty of the detailed biological mechanisms driving the respective networks in cells, we will focus our analysis on the established generic features in network biology which provide a reasonable approximation of the reality of states in single cells which follows a log-normal distribution. This implies systems where any entity can exhibit only two states of activity (boolean states - active and inactive \ie\ $0$ and $1$) \citep{razquin_2018}. 

To our knowledge, such an analysis of an Ising model with asymmetric states of activity has not been investigated so far. 
The objective of this paper is to establish a numerical and theoretical framework for such a \textit{\mim} for a chosen simulated scale-free network \ie\ \banet\ whose degree distribution closely resembles that of real-world biological systems. Preliminary results of this work have been presented in the form of a poster and talk \citep{krishnan_2018, krishnan_2019}. We study the conditions under which this network of \textit{modified Ising spins or gene-type spins} undergoes phase transition under the influence of temperature and magnetic field. The paper is organized as follows: \sect\ \ref{sec:background} provides a short overview of the Ising model and terminologies used in the subsequent sections of the paper; in \sect\ \ref{sec:numerical} we show the conditions under which the \mim\ can undergo phase transitions for different initial configurations of the system (for positive and negative coupling constants) using Monte Carlo simulations; \sect\ \ref{sec:analytical_methods} presents the mean-field solutions and shows a mapping between classical Ising model and the \mim.
\section{Background}
\label{sec:background}

The Hamiltonian of the Ising model of a network reads,

\begin{align}
H = - \frac{1}{2} \sum_{ij}\cm \si\sj - \hh \sum_{i} \si  \hspace{1cm} \cm = \cc\adj
\label{eq:hamiltonian}
\end{align}

\noindent where $\cc$ is the coupling constant specifying the strength of interactions; $\adj$ is the adjacency matrix; $\hh$ indicates a constant external field; $\cm\si\sj$ is the coupling energy arising due to the interaction between nodes and shows the effect of cooperative behavior; $\hh \sum_{i} \si$ is the energy arising due to the effect of magnetic field. The Hamiltonian so formed from these two terms is the total energy of the system. If $\cc>0$, neighboring spins prefer to take the same values (referred to as ferromagnetic exchange interaction in a classical Ising model); when $\cc<0$ neighboring spins prefer to take opposite values (referred to as anti-ferromagnetic exchange interaction in a classical Ising model). Spins, $\si, \sj$ can take values $-1$ and $1$ in the classical Ising model; and $0$ and $1$ in the \mim. 

We study the system in the canonical ensemble wherein we keep the temperature as the controlling parameter of the system allowing energy to change. The probability of finding a particular spin configuration $\si$ is, $\frac{1}{\z} \mathrm{exp}(-\beta H({s_i}))$ where $\z$ is the partition function wherein the negative sign indicates higher probability for lower energy states; and $\beta = \frac{1}{\kb \tp}$ is the inverse temperature that cancels whatever dimensions the Hamiltonian may have ($\kb$ assumed to be equal to $1$ in this paper). From the partition function, we can calculate the thermodynamic properties of the system such as magnetization, internal energy, etc. The order parameter is defined as,

\begin{equation}
\mg = \frac{1}{\nn}\sum_{i} \si
\label{eq:mag}
\end{equation}

A sudden change occurring in the behavior of magnetization (or order parameter) with respect to temperature indicates phase transition from ferromagnetic to paramagnetic state in the system. Phase transitions in which discontinuity occurs in order parameter (in the first derivative of the free energy) is referred to as first-order phase transitions; continuous in order parameter is referred to as second-order phase transitions. The temperature at which such a transition occurs is called the critical temperature, $\tc$. The Hamiltonian in \eq\ \ref{eq:hamiltonian} has been treated for the case of classical Ising model of a ferromagnetically-coupled \banet\ analytically by \cite{bianconi_2002} and numerically by \cite{aleksiejuk_2002}. For ferromagnetic to the paramagnetic phase transition, such a system has infinite critical temperature $T_c$ and the effective critical temperature increases as the logarithm of the system size.

Solving such a system is a computationally intensive problem, and solving exactly analytically is difficult as well. We, therefore, use Markov Chain Monte Carlo (MCMC) methods to randomly change the state of the system and accept it or reject it according to a given probability function until the system achieves thermodynamic equilibrium. Specifically, we use the Metropolis algorithm \citep{metropolis_1953}, which is a type of MC method implemented as follows: a network node is chosen randomly, and its spin is noted. The cost of switching this state is calculated as the energy difference between its current state and flipped state. If this cost is negative, the flip is accepted. Else, a random number drawn from a uniform distribution is generated. If this is smaller than energy difference, the flip is accepted. Else the current spin state of the node is preserved. 

This algorithm changes as a single spin per iteration, which means that the program will explore the state space very slowly and therefore will need a considerable number of iterations to get a good approximation of the partition function. At every temperature, the system will move very slowly towards the part of the state space that corresponds to that temperature, thereby taking time to reach thermal equilibrium.

\section{Numerical Simulations}
\label{sec:numerical}

As motivated in \sect\ \ref{sec:intro} the focus of this paper is to study the system in \eq\ \ref{eq:hamiltonian} for modified Ising spins of the \banet. Such a network is constructed based on two main properties of a real-world network - linear growth and preferential attachment\citep{albert_2002}. The network is initialized with $\mm_0$ nodes that are not connected. Subsequently, new nodes with $\mm$ edges are added in iterations to the existing $\mm_0$ nodes. The resultant network has a power-law degree distribution and is characterized by a degree exponent, $2 < \gm < 3$ that resembles real-world biological networks (\cf\ \app\ \ref{sec:appdx}). 

We now put modified Ising spins on nodes of a \banet\ of size, $\nn = 5 \times 10^{3}$ and preferentially attached links, $\mm = 5$. Then with the standard heat bath Monte Carlo algorithm, we do a spin search for thermal equilibrium at temperature $\tp$. We equilibrate the system for $2 \times 10^4$ MC steps. After this transient period, we simulate $3 \times 10^4$ MC steps which allow for an average $10$ spin flips per spin and then sample at the end of every step. We perform simulations for both ferromagnetically and anti-ferromagnetically coupled networks, under the influence and absence of the magnetic field.

\begin{figure}[!htb]
\includegraphics[scale = 0.5]{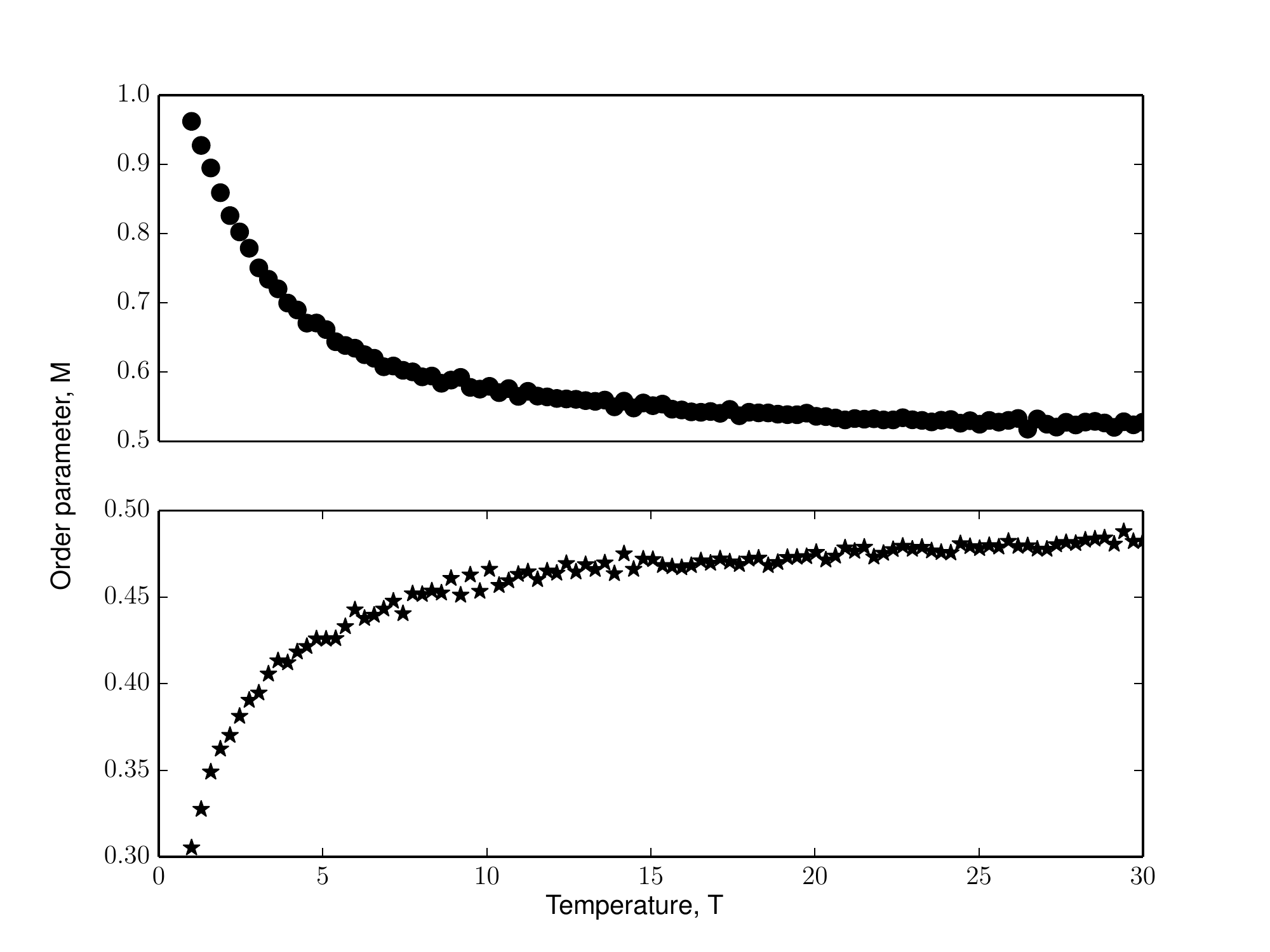}
\caption{Monte Carlo simulations of the \mim\ of a \banet\ at magnetic field, $\hh = 0$. Figure shows evolution of order parameter, $\mg$ as a function of Temperature, $\tp$ . Top panel: \mim\ of \banet\ with positive coupling constant, $\cc$ (indicated by black dots). Bottom panel: \mim\ of \banet\ with negative coupling constant, $- \cc$  (indicated by black stars). Simulation parameters: network size, $\nn = 5 \times 10^3$, preferentially-attached links to construct \banet\ $\mm = 5$, magnitude of coupling constant, $|\cc| = 1$.}
\label{fig:h0}
\end{figure}

Under no influence of the magnetic field and ferromagnetic exchange interaction, all nodes in the network start at an active state where the order parameter, $\mg = 1$. At $\tp < 1$, the system favors order as seen in the top panel of \fig \ref{fig:h0}. As the thermal fluctuations in the system increases, the disorder in the system increases. The order parameter reaches $\frac{1}{2}$ asymptotically as $T \rightarrow \infty$. Similarly, when the system is initialized with an anti-ferromagnetic exchange interaction, we see that all nodes start at an inactive state as seen in the bottom panel of \fig\ \ref{fig:h0} at $\tp <1$. As thermal fluctuations increases, the order parameter asymptotically reaches $\frac{1}{2}$. 

\begin{figure}[!htb]
\textbf{(A)}\includegraphics[scale = 0.35]{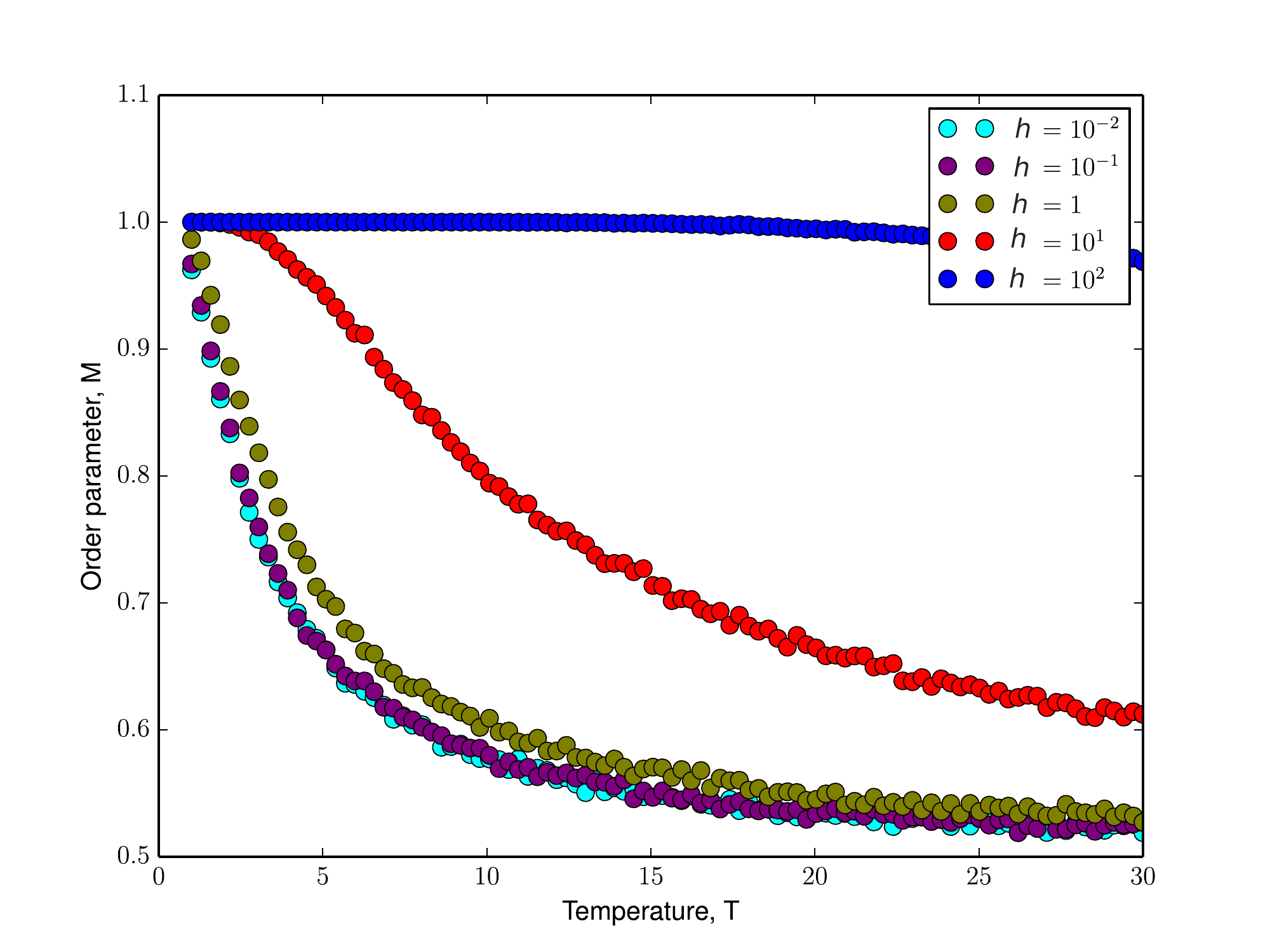}
\textbf{(B)}\includegraphics[scale = 0.35]{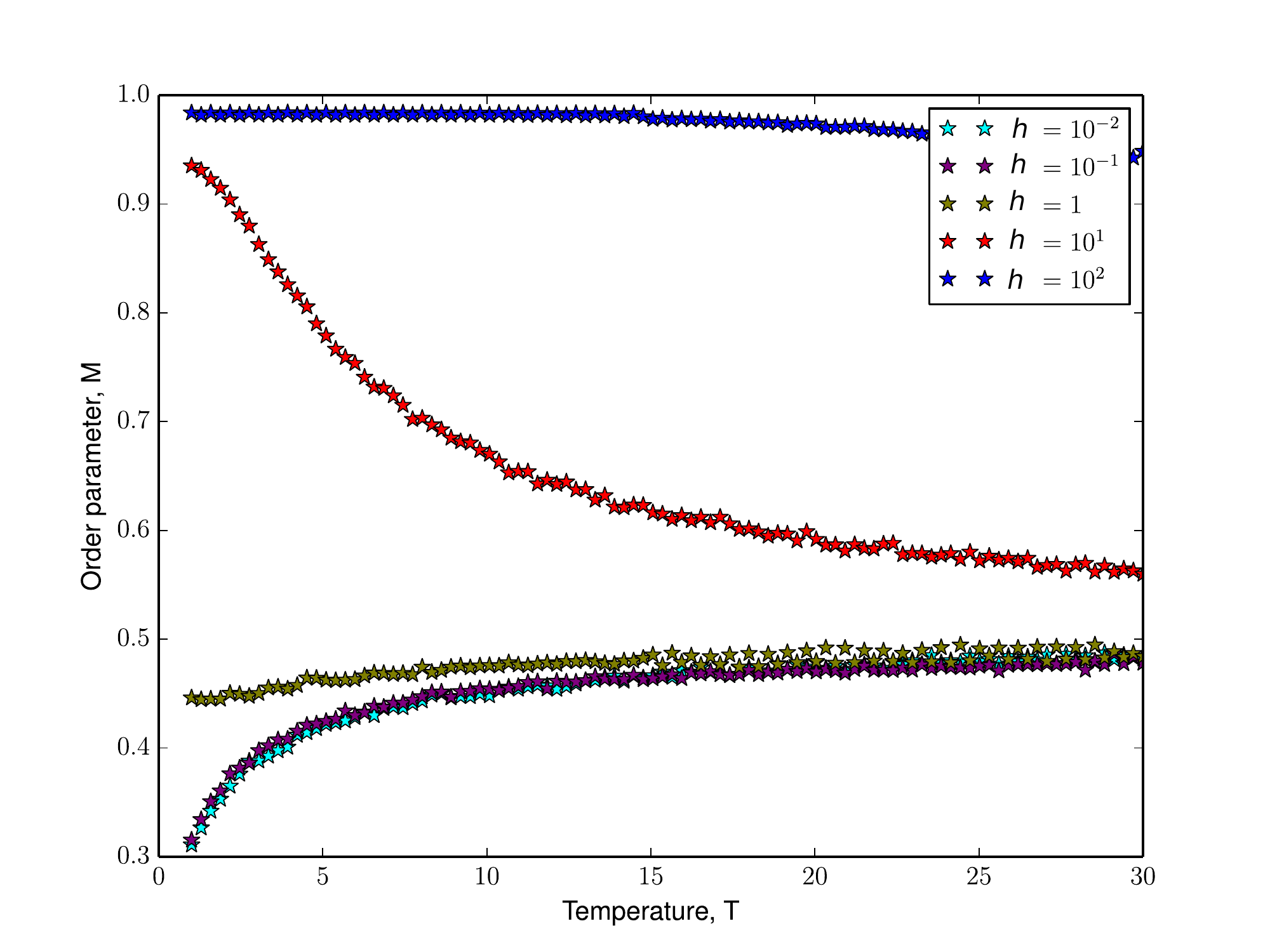}\\
\caption{Monte Carlo simulations of the \mim\ of a \banet\ in the presence of a positive magnetic field, $\hh > 0$ of different magnitudes. Figure shows evolution of order parameter, $\mg$ as a function of Temperature, $\tp$. \textbf{(A)} \mim\ of \banet\ with positive coupling constant, $\cc$ (indicated by dots). \textbf{(B)} \mim\ of \banet\ with negative coupling constant, $- \cc$ (indicated by stars). Simulation parameters: network size, $\nn = 5 \times 10^3$, preferentially-attached links to construct \banet, $\mm = 5$ and magnitude of coupling constant, $|\cc| = 1$.}
\label{fig:hg0}
\end{figure}

\begin{figure}[!htb]
\textbf{(A)}\includegraphics[scale = 0.35]{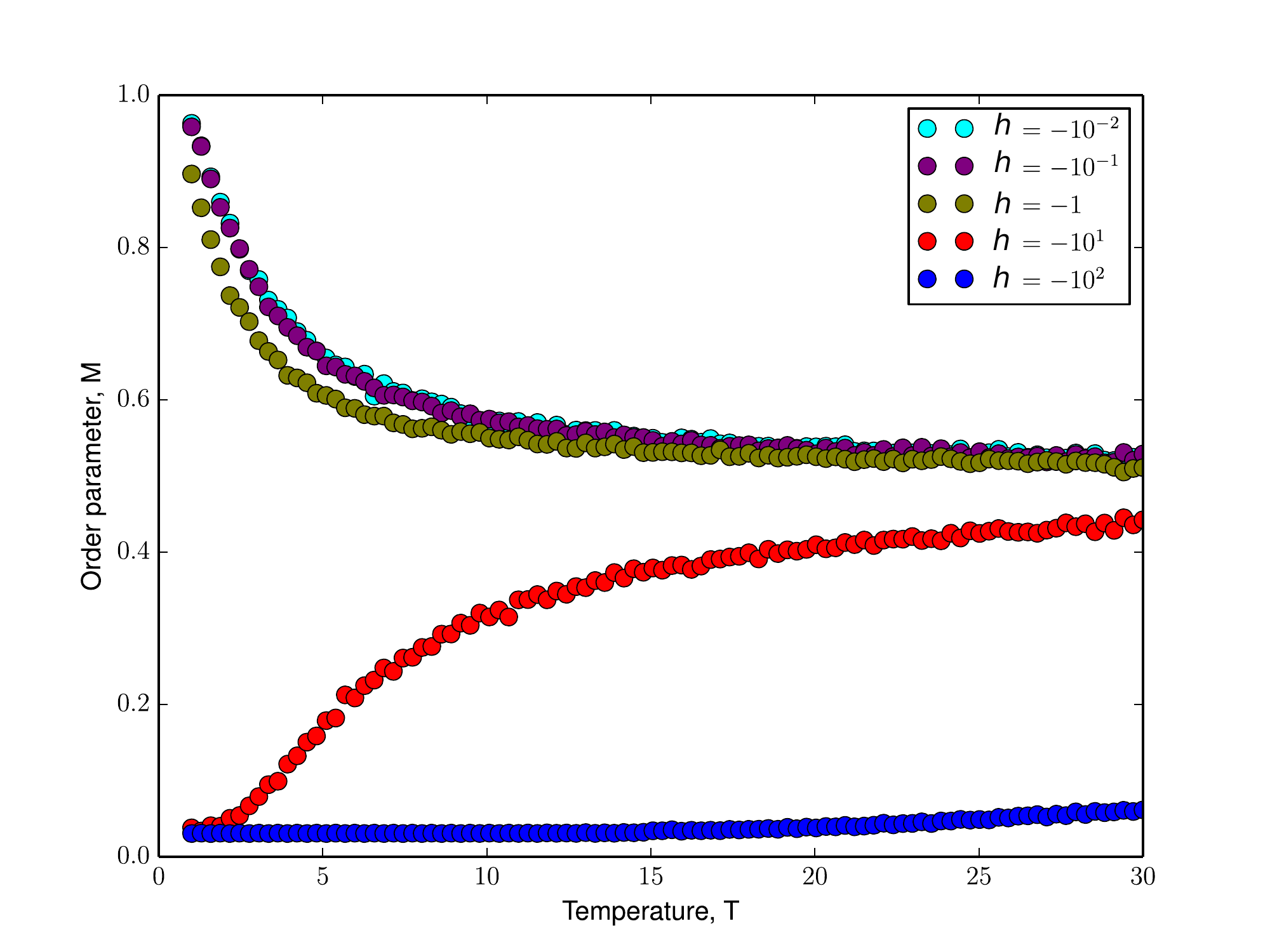}
\textbf{(B)}\includegraphics[scale = 0.35]{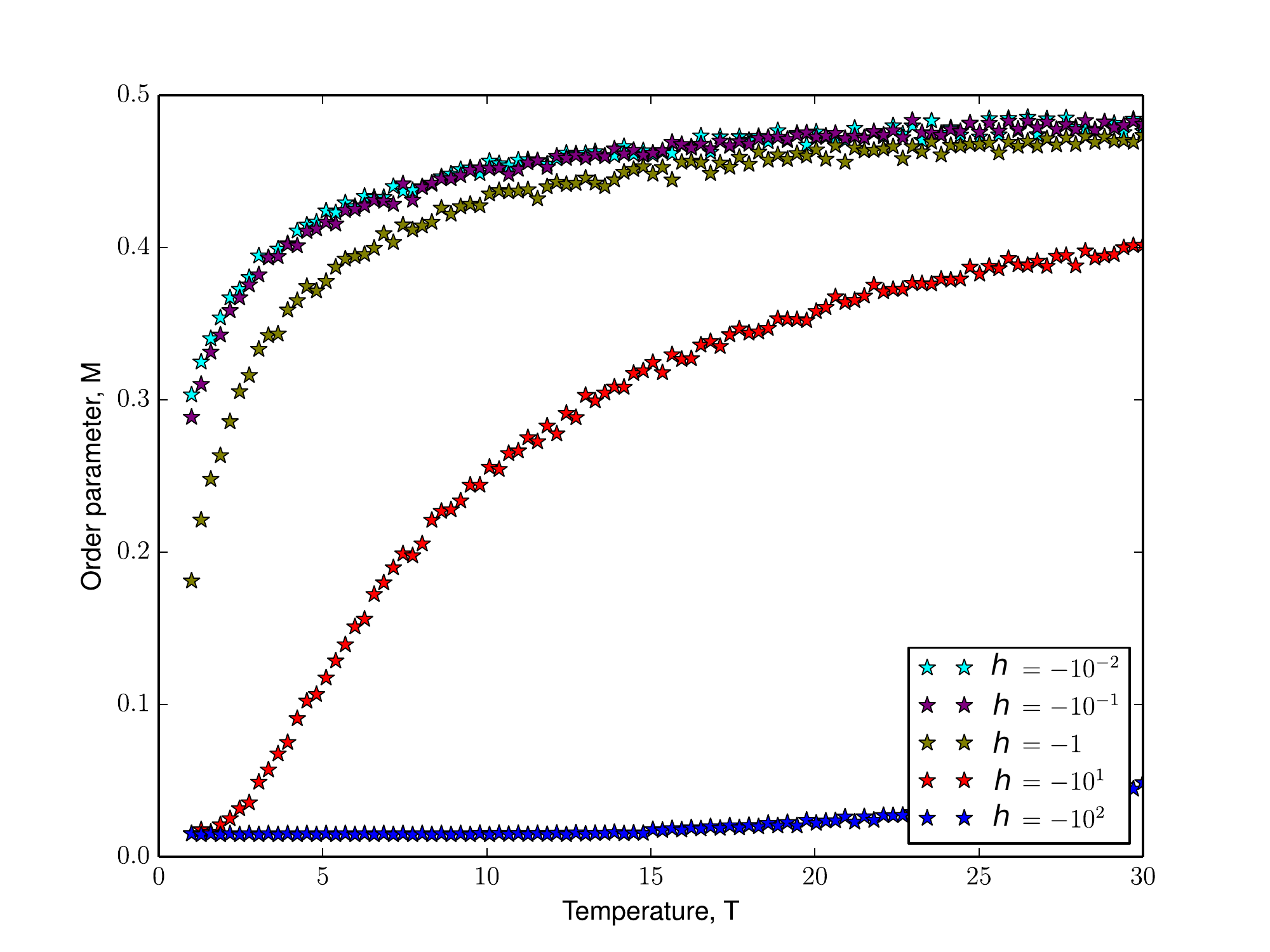}\\
\caption{Monte Carlo simulations of the \mim\ of a \banet\ in the presence of a negative magnetic field, $\hh < 0$ of different magnitudes. Figure shows evolution of order parameter, $M$ as a function of Temperature, $\tp$. \textbf{(A)} \mim\ of \banet\  with positive coupling constant, $\cc$ (indicated by dots). \textbf{(B)} \mim\ of \banet\ with negative coupling constant, $- \cc$ (indicated by stars). Simulation parameters: network size, $\nn =  5 \times 10^3$, preferentially-attached links to construct \banet\ $\mm = 5$, magnitude of coupling constant, $|\cc| = 1$.}
\label{fig:hl0}
\end{figure}

Under the influence of magnetic field, the system behavior changes as seen in \figs\ \ref{fig:hg0} and \ref{fig:hl0}. Consider the ferromagnetically coupled \mim\ of \banet\ influenced by positive magnetic field (\fig\ \ref{fig:hg0}(A)). The field term in the Hamiltonian is effectively a constant holding the network above the mean of two states at $\frac{1}{2}$. The larger the magnitude of the magnetic field, the more unlikely it is to induce disorder in the network. For an anti-ferromagnetically coupled \mim\ of \banet, we observe for that for small magnitudes of the positive magnetic field ($\hh < 1$), the asymptotic property of order parameter vanishes as in the case of a ferromagnetically-coupled system (\fig\ \ref{fig:hg0}(B)). 

\begin{figure}[!htb]
\includegraphics[scale = 0.4]{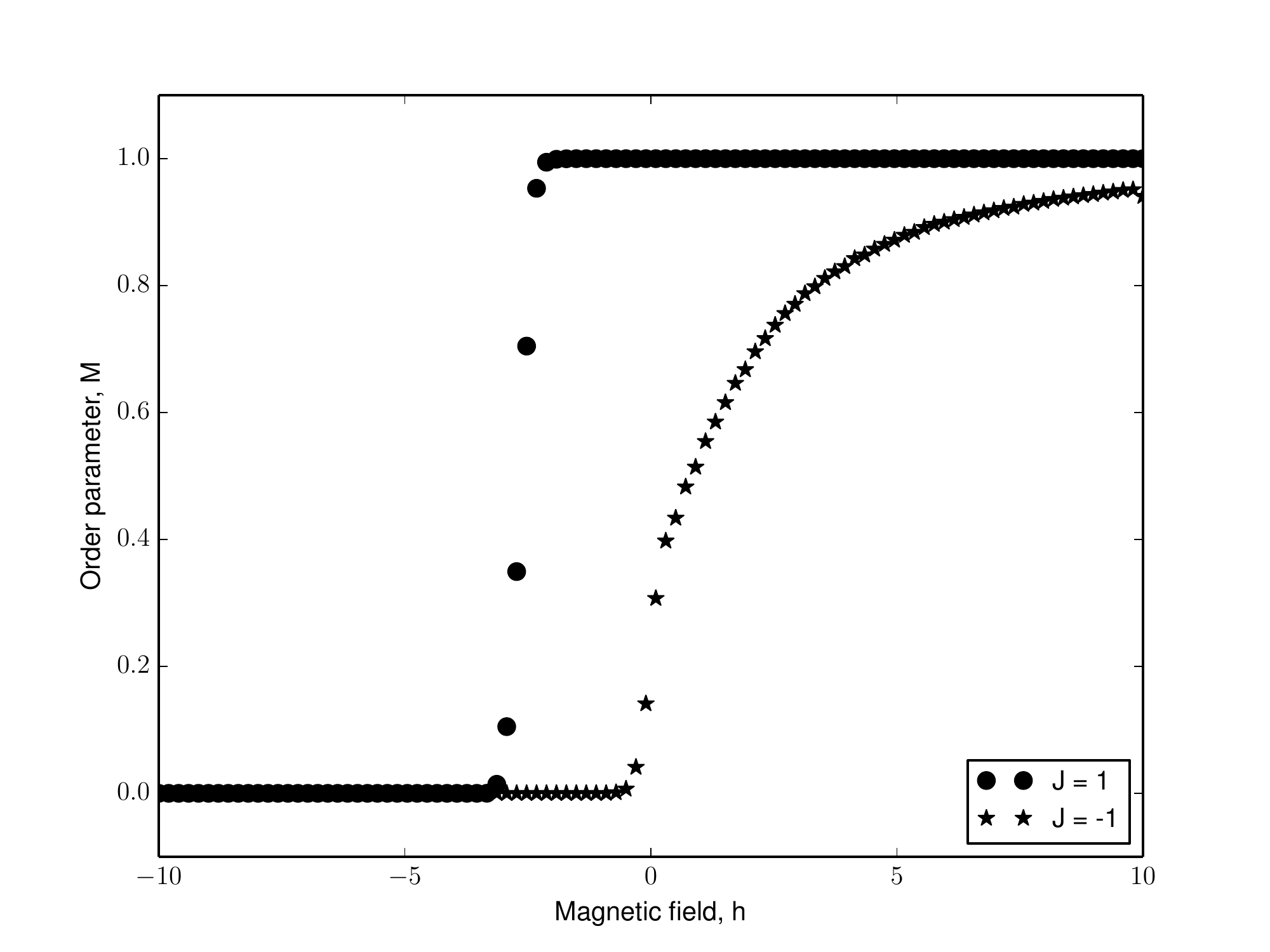}
\caption{The \mim\ of a \banet\ exhibits phase transition under the influence of magnetic field at a fixed Temperature, $\tp = 0.1$. Black dots indicate the order parameter trend for a \mim\ of \banet\  with positive coupling constant, $\cc = 1$. Black stars indicate the order parameter trend for a \mim\ of \banet\ with negative coupling constant, $\cc = -1$.}
\label{fig:banetpt}
\end{figure}

However, for higher magnitudes of the magnetic field, we observe that the field term can trigger activity in the network \ie\ switch from $\mg = 0$ to $\mg = 1$ at $0 < \tp < 1$ and subsequently follow the dynamics of a ferromagnetically-coupled system. A negative magnetic field, on the other hand, inverts the dynamics of a ferromagnetically-coupled \mim\ instead. As can be seen in \fig\ \ref{fig:hl0}(A), at $-2.5<\hh<0$ there is an abrupt drop in the order parameter to $0$ and for lower values the network remains inactive (as can be verified from our observations in \figs\ \ref{fig:hg0} and \ref{fig:hl0}). An anti-ferromagnetically coupled network has order parameter $\mg=0$ at $\hh=0$. Lower values of the magnetic field keep the network in the inactive state. For a positive magnetic field, the network undergoes a relatively smooth (almost abrupt) phase transition to the active state. Owing to this, unlike in a ferromagnetically coupled network, we observe intermediate values of order parameter and $\mg \rightarrow 1$ as $\hh$ increases, confirming our observations in \figs\ \ref{fig:hg0} and \ref{fig:hl0}. Thus we can infer that the \mim\ of a \banet\ undergoes phase transition due to the magnetic field as shown in \fig\ \ref{fig:banetpt}. 

We see that the transition has a discontinuity in order parameter and hence this may be a first-order phase transition. Systems that undergo first-order phase transition are characterized by hysteresis loops. This implies that the network may show more than one value of order parameter for a given magnetic field, $\hh$. The hysteresis loop shows the dependence of the state of the system on its history, and it is this phenomenon that forms memory in a hard disk drive. 

The procedure to investigate the existence of hysteresis has been well-established, particularly in the context of magnetic materials. We apply the same method for the \mim\ of a \banet\ summarized shortly here. Starting with a high negative magnetic field, $\hh$, and a stable configuration of the system, we increase the field slowly. For some value of $\hh$, the local field for a node flips. This causes changes in the effective field of the nodes connected to this node, thereby causing them to flip. Once the flipping in the system has thermalized, the order parameter of the system is measured. Subsequently, the magnetic field is increased slightly, and the process repeated until the order parameter attains a stable state. This way, one can obtain one half of the hysteresis loop (for $\hh$ from $-\infty$ to $\infty$). The other half of the hysteresis loop is obtained when the magnetic field, $\hh$ is decreased (for $\hh$ from $\infty$ to $-\infty$).

\begin{figure}[!htb]
\textbf{(A)}\includegraphics[scale = 0.35]{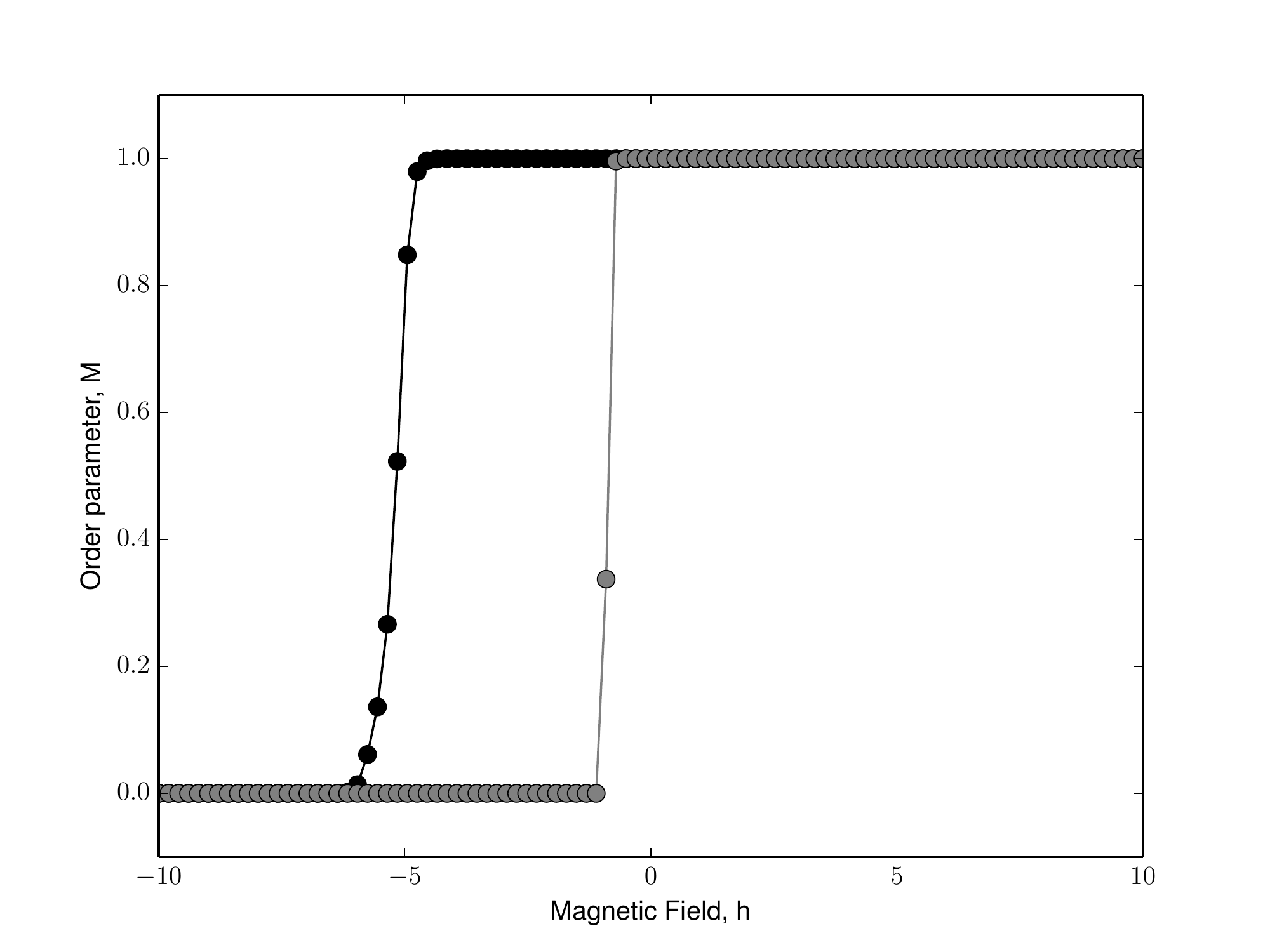}
\textbf{(B)}\includegraphics[scale = 0.35]{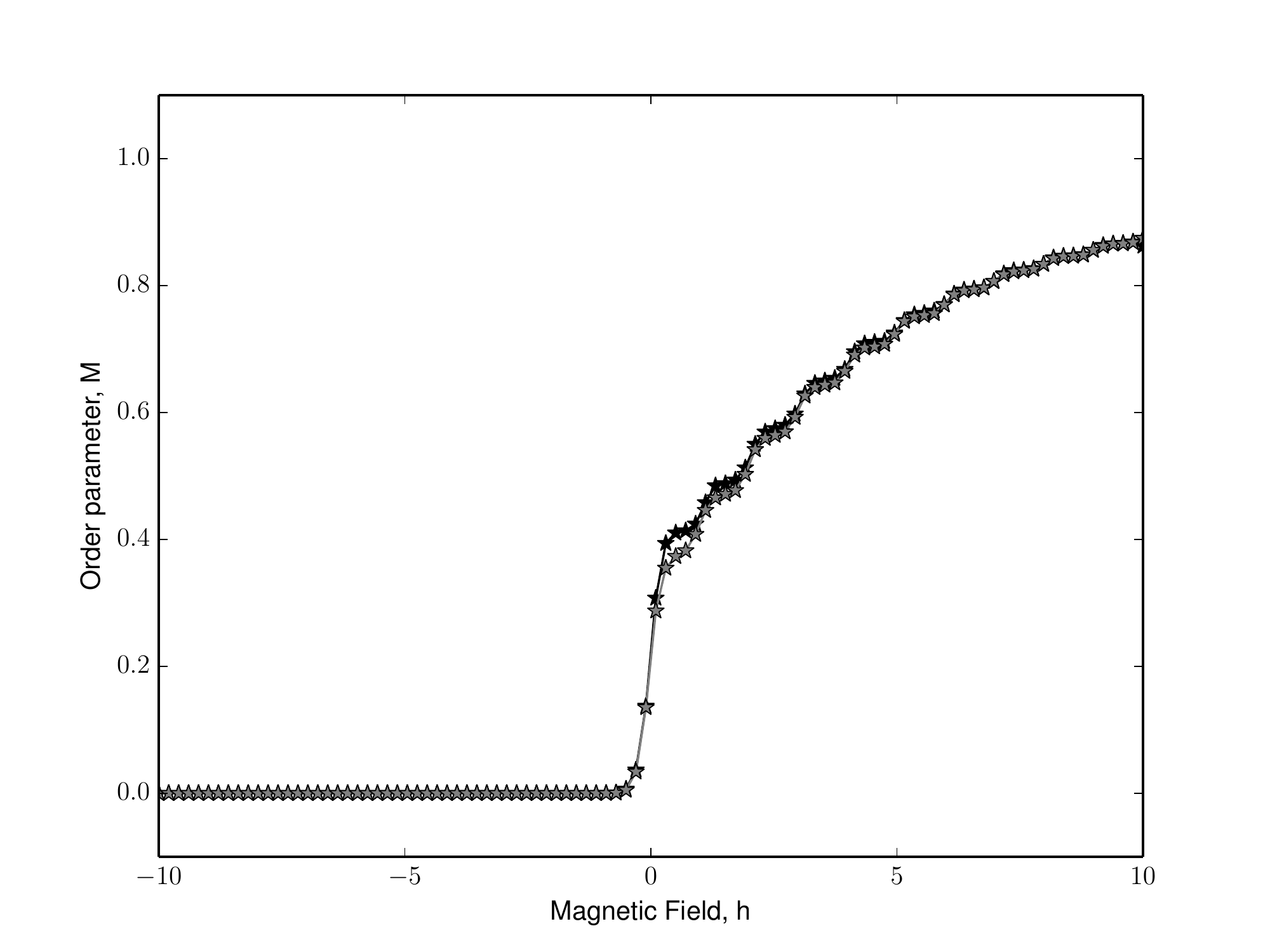}\\
\caption{The \mim\ of a \banet\ exhibits hysteresis: \textbf{(A)} ferromagnetically coupled, $\cc = 2$ (indicated by dots). \textbf{(B)} anti-ferromagnetically coupled, $\cc = -2$ respectively (indicated by stars). Simulation parameters: $\nn = 5 \times 10^3$ and preferentially attached links, $\mm = 5$. The gray curve indicates order parameter as we drive the system forward from $\hh_0 = 10$ to $\hh_n = -10$ and the black curve as we drive the system backward from $\hh_0 = -10$ to $\hh_n = 10$.}
\label{fig:hyst}
\end{figure}

A typical hysteresis loop takes the form of a sigmoid, however, in the case of a ferromagnetically coupled \mim\ the loop is almost a rectangle as can be seen in \fig\ \ref{fig:hyst}(A). An anti-ferromagnetically coupled network does not exhibit hysteresis for low coupling constants as can be seen in \fig\ \ref{fig:hyst}(B). We will analyze these observations and discuss the asymptotic behavior in detail using analytical approaches in \sect\ \ref{sec:analytical_methods}.

\clearpage
\section{Analytical Methods}
\label{sec:analytical_methods}
\subsection{Mean Field Approximation}
\label{subsec:mft}

One of the most important analytical tool to study disordered systems is represented by mean-field theories. Mean field theory is frequently used due to its conceptual simplicity, as a useful tool, especially when there is no exact solution for the problem. This approximation is used to reduce an interacting problem to a non-interacting one which is easier to solve. Let us consider the \mim\ of a \banet\ treated numerically in \sect\ \ref{sec:numerical}. Rewriting the Hamiltonian of the ferromagnetically-coupled system with gene-type spins, $\gs$,

\begin{equation}
\hm_{\gs} = -\frac{1}{2}\sum_{i,j=1}^{\nn}\cm\si\sj - \hh \sum_{i=1}^{\nn}\si \hspace{0.5cm} \si = \gs \hspace{0.3cm} \cc > 0
\label{eq:hamiltonian}
\end{equation}

\noindent where $\cm = \cc \adj$. Since the adjacency matrix $\adj$ is symmetric, the factor $\frac{1}{2}$ is included so as not to count any pairs twice. We can write the interactions between neighboring spins in terms of their deviations from the average spin $\mg$ as,

\begin{equation}
\begin{split}
\si\sj &= [(\si - \mg) + \mg][(\sj - \mg) + \mg]\\
&= (\si - \mg)(\sj - \mg) + \mg(\sj - \mg) + \mg(\si - \mg)+ \mg^2\\
\label{eq:avgspin}
\end{split}
\end{equation}

\noindent where $\mg = \frac{1}{N}\sum_{i=1}^{\nn}\si$ is the order parameter. Assuming that the fluctuations around the mean spin is small, the Hamiltonian can be rewritten as,

\begin{equation}
\begin{split}
\hmf &= -\frac{1}{2} \sum_{i,j=1}^{\nn}\cm 
[\mg(\sj - \mg) + \mg(\si - \mg) + \mg^2] - \hh \sum_{i=1}^{\nn}\si\\
&= -\Big[ \frac{\cc\mm}{2} \sum_{i=1}^{\nn} \sum_{j=1}^{\nn} \adj \si + \frac{\cc \mg}{2} \sum_{i=1}^{\nn}\sum_{j=1}^{\nn}\adj\sj - \frac{\cc \mg^2}{2} \sum_{i,j=1}^{\nn} \adj \Big] -\hh \sum_{i=1}^{\nn}\si\\
\label{eq:hmf}
\end{split}
\end{equation}

\noindent Consider the second term in the right hand side of \eq\ \ref{eq:hmf}. This can be written as $(i \rightarrow j)$:

\begin{equation}
\frac{\cc \mg}{2} \sum_{i=1}^{\nn}\sum_{j=1}^{\nn}A_{ji}\si = \frac{\cc\mg}{2}\sum_{i=1}^{\nn}\sum_{j=1}^{\nn}\adj \sj
\label{eq:sym}
\end{equation}

\noindent since $\adj = A_{ji}$, $A$ is symmetric. Therefore from \eqs\ \ref{eq:hmf} and \ref{eq:sym}, 

\begin{equation}
\hmf = \frac{\cc \mg^2}{2}\sum_{i,j}^{\nn}\adj - \cc \mg \sum_{i,j}^{\nn} \adj \si - \hh \sum_{i=1}^{\nn}\si
\label{eq:hmfr}
\end{equation}

\noindent This is the mean-field Hamiltonian for a chosen realization of the network. So the ensemble average of the Hamiltonian of the system is,

\begin{equation}
\langle \hmf \rangle = \frac{\cc \mg^2}{2} \sum_{i,j}^{\nn} \langle \adj \rangle - \cc \mg \sum_{i,j}^{\nn}\langle \adj \rangle \si - \hh \sum_{i=1}^{\nn}\si
\label{eq:hmfavg1}
\end{equation}

\noindent For a \banet, 

\begin{equation}
\langle \adj \rangle = \pij = \frac{1}{2\mm\nn}\kk_{i}\kk_{j}
\label{eq:adjavg}
\end{equation}

\noindent where $\kk_{i}$ is the number of links of the $i$th node of the network \cite{bianconi_2002} (cf. \app\ \ref{sec:appdx}). From \eqs \ref{eq:hmfavg1} and \ref{eq:adjavg}, using the relation $\sum_{i=1}^{\nn}\kk_{i} = \sum_{j=1}^{\nn} \approx 2\mm\nn$,

\begin{equation}
\begin{split}
\langle \hmf \rangle &= \frac{\cc \mg^2}{2}\sum_{i,j=1}^{\nn} \frac{1}{2\mm\nn}\kk_{i}\kk_{j} - \cc \mm \sum_{i,j=1}^{\nn} \frac{1}{2\mm\nn}\kk_{i}\kk_{j}\si - \hh \sum_{i=1}^{\nn}\si\\
&= \frac{\cc \mg^2}{4\mm\nn}\sum_{i=1}^{\nn}\kk_{i}\sum_{j=1}^{\nn}\kk_{j} - \frac{\cc\mg}{2\mm\nn}\sum_{j=1}^{\nn}\kk_{j}\sum_{i=1}^{\nn}\kk_{i}\si - \hh \sum_{i=1}^{\nn}\si\\
&= \frac{\cc\mg^2}{4\mm\nn} \times 2\mm\nn \times 2\mm\nn - \cc \mg \sum_{i=1}^{\nn} \kk_{i}\si - \hh \sum_{i=1}^{\nn}\si\\
&= \cc \mg^2 \mm \nn - \underbrace{(\hh + \cc \mm \kk_{i})}_{\heff_{i}}\si\\
\langle \hmf \rangle &= \cc\mg^2\mm\nn - \sum_{i=1}^{\nn} \heff_{i}\si, \hspace{1cm} \heff_{i} = (\hh + \cc\mm\kk_{i})\\
\label{eq:heff}
\end{split}
\end{equation}

\noindent Hence the \mim\ of a \banet\ reduces to a system of non-interacting spins in an effective local field, $\heff_{i} = (\hh + \cc\mm\kk_{i})$. The partition function can be evaluated as,

\begin{equation}
\begin{split}
\z &= \sum_{\mathrm{config}} e^{-\bt \langle \hmf \rangle}\\
&= \sum_{\si = \gs} \ldots \sum_{s_{\nn} = \gs} e^{-\bt \Big[ \cc\mg^2\mm\nn - \sum_{i=1}^{\nn} \heff_{i}\si \Big]}\\
&= e^{-\bt \cc \mg^2 \mm \nn} \prod_{i} \Big( \sum_{\gs} e^{\bt \heff_{i}}\si\Big)\\
\z &= e^{-\bt \cc \mg^2 \mm \nn} \prod_{i} \Big( 1 + e^{\bt \heff_{i}}\Big) \\
\label{eq:ptfunc}
\end{split}
\end{equation}

\noindent The mean spin, $\mg$ can be calculated from the partition function using the following relation:

\begin{equation}
\begin{split}
\mg &=\frac{1}{\nn} \sum_{i=1}^{\nn}\si\\
&= \frac{1}{\nn\bt} \frac{\partial \ln\z}{\partial \hh}\\
\label{eq:mspin1}
\end{split}
\end{equation}

\noindent From this, evaluating $\ln\z$,

\begin{equation}
\begin{split}
\ln\z &= - \bt\cc\mg^2\mm\nn + \ln \prod_{i}\Big[ 1 + e^{\bt\big({\hh + \cc\mg\kk_{i}}\big)}\Big]\\
&= - \bt\cc\mg^2\mm\nn + \sum_{i} \ln \Big[ 1 + e^{\bt(\hh + \cc \mg \kk_{i})}\Big]\\
\label{eq:lnz}
\end{split}
\end{equation}

\noindent Therefore from \eqs\ \ref{eq:mspin1} and \ref{eq:lnz},

\begin{equation}
\begin{split}
\mg &= \frac{1}{\nn\bt} \sum_{i=1}^{\nn} \frac{\bt e^{\bt(\hh + \cc\mg\kk_{i})}}{1 + e^{\bt(\hh+\cc\mg\kk_{i})}}\\
&= \frac{1}{\nn} \sum_{i=1}^{\nn}\frac{e^{\bt(\hh + \cc\mg\kk_{i})}}{1 + e^{\bt(\hh + \cc\mg\kk_{i})}}
\label{eq:mspin2}
\end{split}
\end{equation}

\noindent Therefore the central mean-field equation for ferromagnetically coupled \banet\ with asymmetric spins takes the implicit form,

\begin{equation}
\mg = \frac{1}{\nn} \sum_{i=1}^{\nn} \frac{1}{1+e^{-\bt(\hh + \cc\mg\kk_{i})}}
\label{eq:mfe1}
\end{equation}

\noindent Similarly for anti-ferromagnetically coupled \banet\ $(\cc \rightarrow -\cc)$ the central mean-field equation is,

\begin{equation}
\mg = \frac{1}{\nn} \sum_{i=1}^{\nn} \frac{1}{1+e^{-\bt(\hh - \cc\mg\kk_{i})}}
\label{eq:mfe2}
\end{equation}

\noindent Note that the order parameter depends on the coupling constant, $\cc$ and node degree, $\kk_{i}$. Let us first study the behavior of the system in the absence of magnetic field. The mean-field equation for ferromagnetically coupled \banet\ with gene-type spins and no external field is,

\begin{equation}
\mg = \frac{1}{\nn}\sum_{i=1}^{\nn} \frac{1}{1+e^{\pm \bt\cc\mg\kk_{i}}}
\label{eq:mfe3}
\end{equation}

\begin{figure}[!htb]
\textbf{(A)}\includegraphics[scale = 0.35]{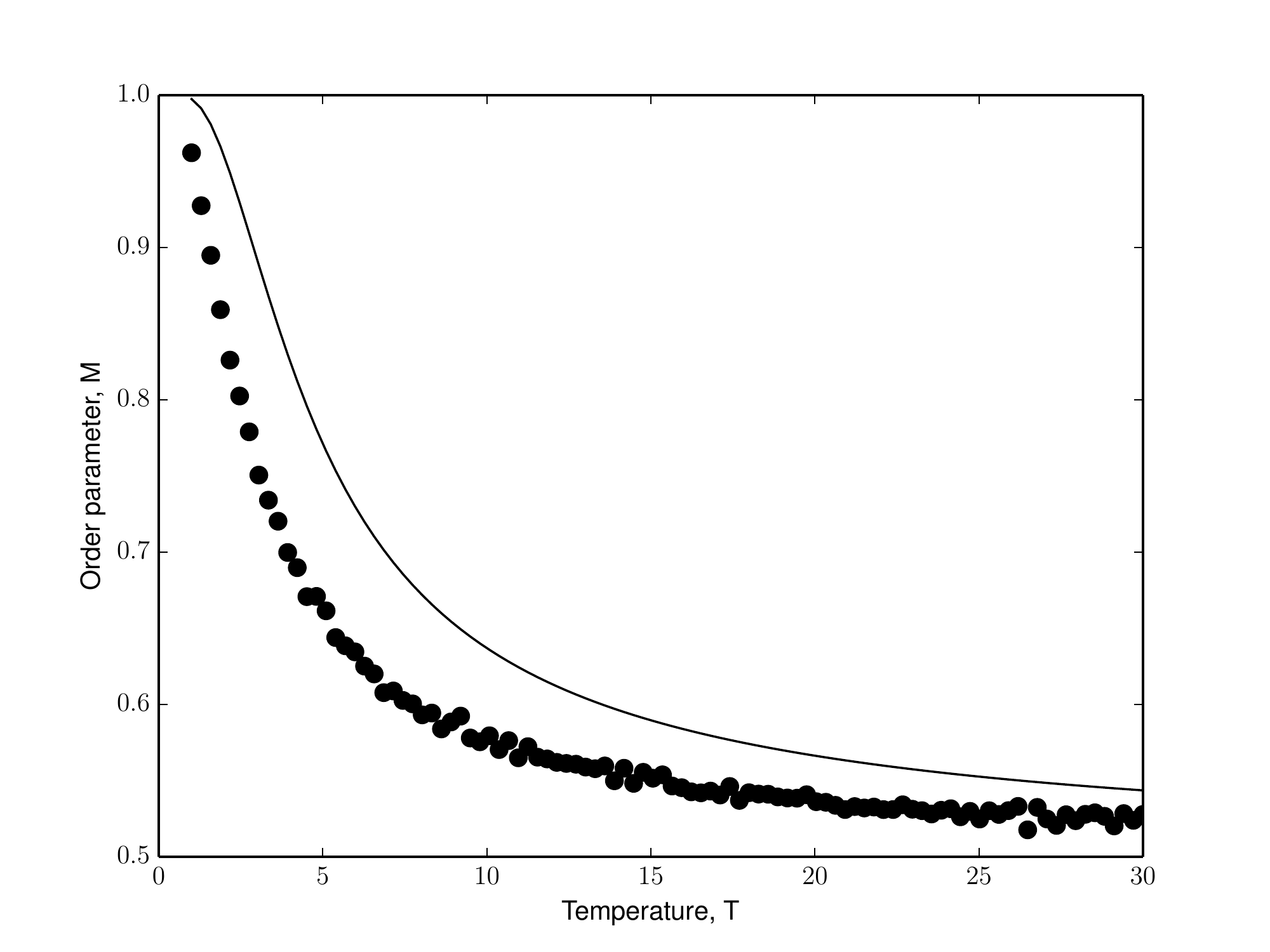}
\textbf{(B)}\includegraphics[scale = 0.35]{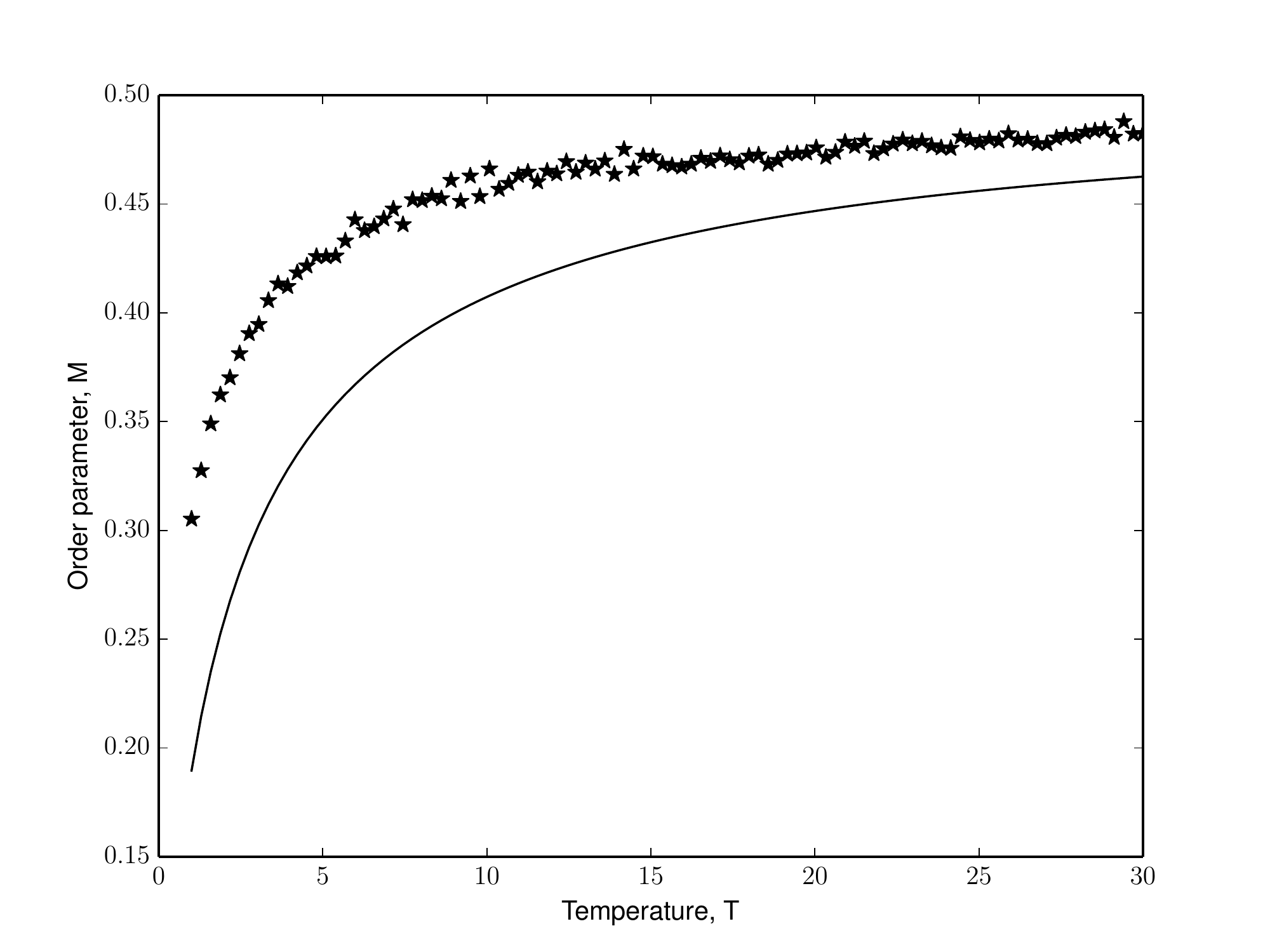}\\
\caption{Mean-Field Theory validates observations from numerical simulations for a \mim\ of a \banet\ in the absence of magnetic field. Figure shows evolution of order parameter, $\mg$ as a function of Temperature, $\tp$: \textbf{(A)} for a \mim\ of \banet\ with positive coupling constant, $\cc$. Black dots indicate Monte Carlo sampling points. Black curve indicates the trend predicted by the central mean-field equation. \textbf{(B)} for a \mim\ of \banet\ with negative coupling constant, $-\cc$. Black stars indicate Monte Carlo sampling points. Black curve indicates the trend predicted by the central mean-field equation. Simulation parameters: network size, $\nn = 5 \times 10^3$, preferentially-attached links to construct \banet\, $\mm = 5$, magnitude of coupling constant, $|\cc| = 1$.}
\label{fig:mft}
\end{figure}

\begin{figure}[!htb]
\textbf{(A)}\includegraphics[scale = 0.35]{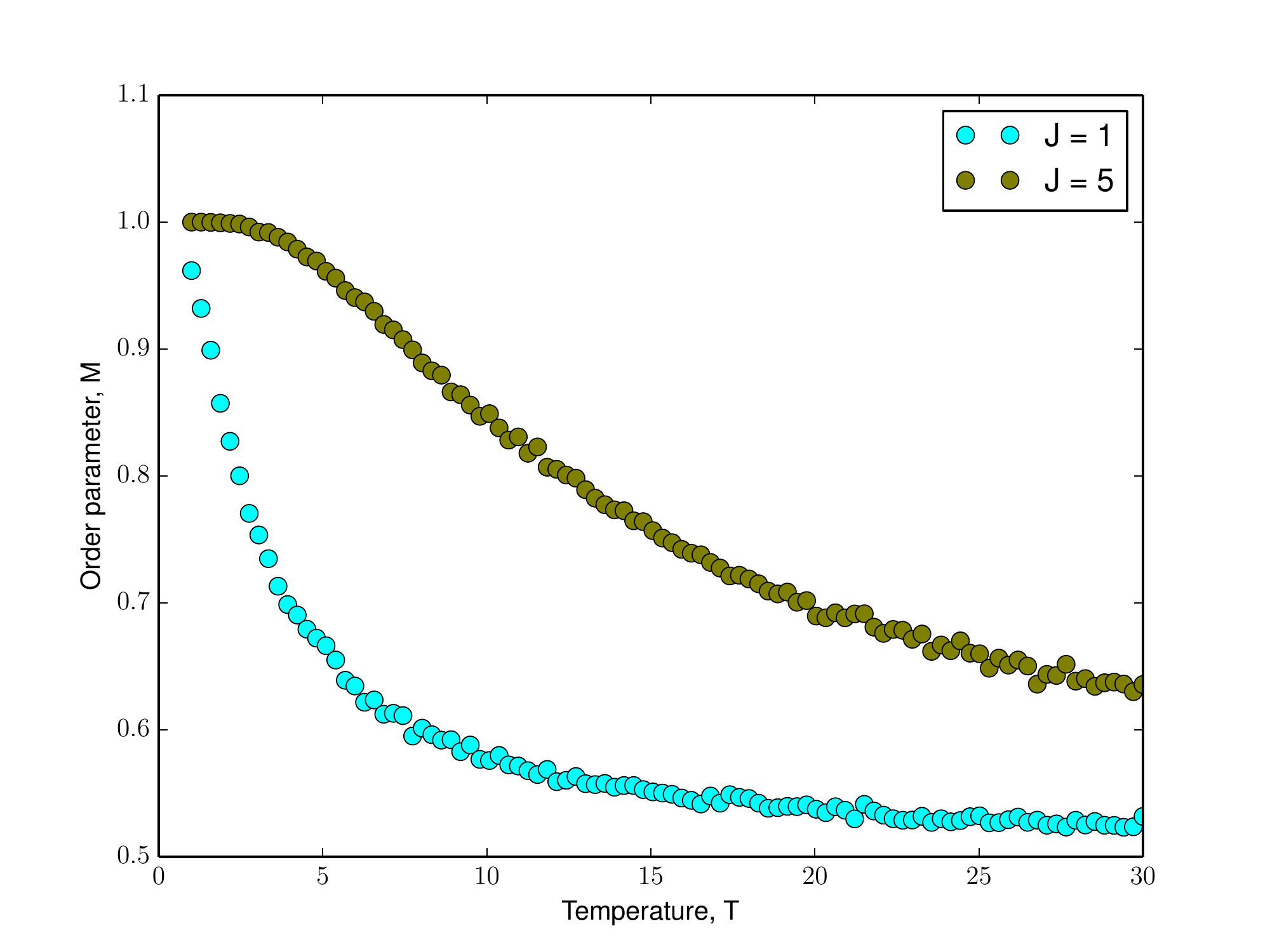}
\textbf{(B)}\includegraphics[scale = 0.35]{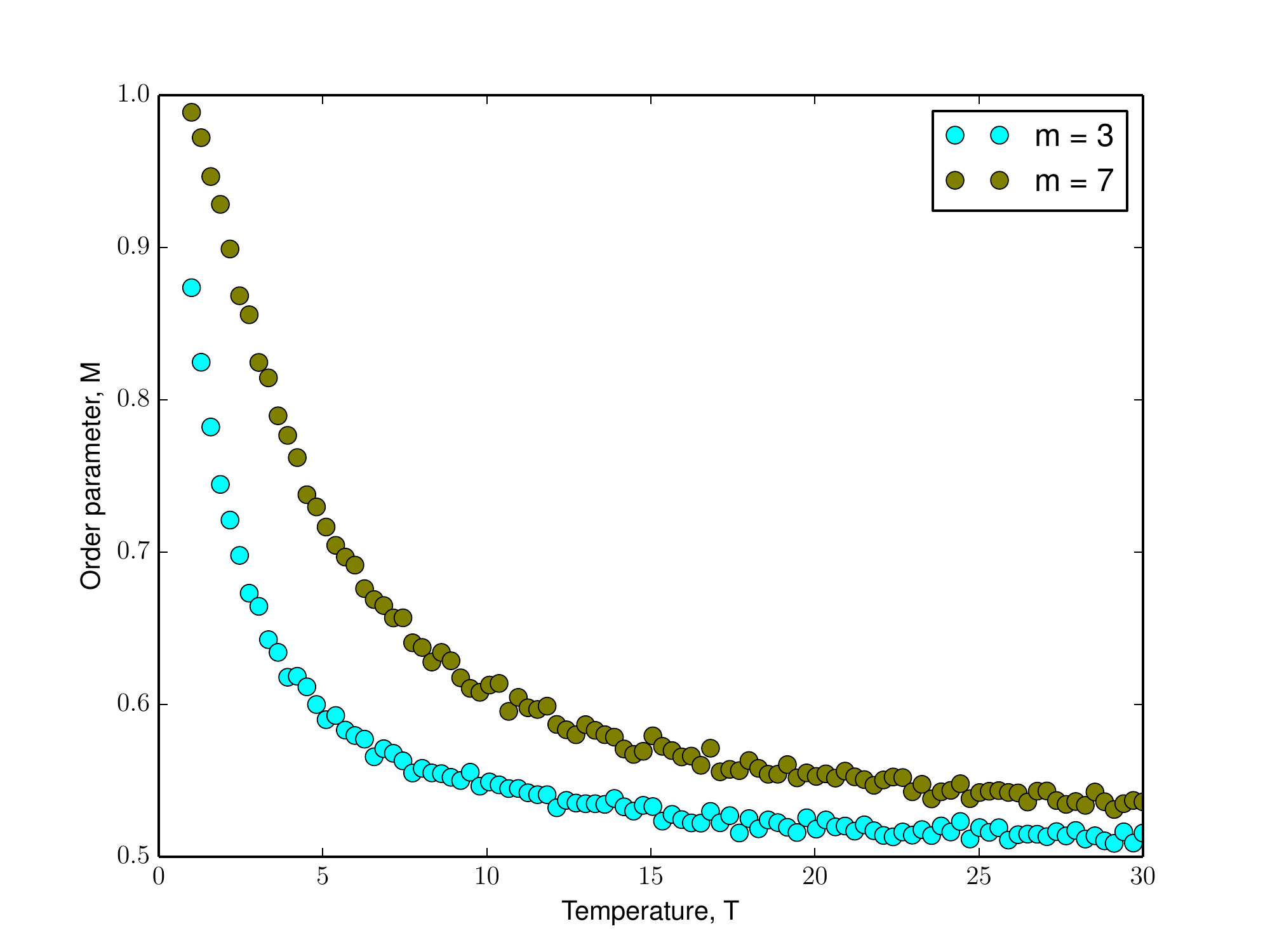}\\
\caption{Monte Carlo simulations of \mim\ of a \banet\ of size, $\nn = 5 \times 10^3$ at $\hh = 0$ and positive coupling constant, $\cc$. \textbf{(A)} for coupling constants, $\cc = 1$ and $\cc = 5$ with $\mm = 3$. \textbf{(B)} for different choice of preferentially attached links, $\mm = 3$ and $\mm = 7$ with $\cc = 1$. Simulation parameters: network size, $\nn = 5 \times 10^3$, preferentially-attached links to construct \banet\ $\mm = 5$, magnitude of coupling constant, $|\cc| = 1$.}
\label{fig:comparison}
\end{figure}

\noindent where $\pm$ stands for ferromagnetically and anti-ferromagnetically coupling respectively. From \eq\ \ref{eq:mfe3} we can investigate the asymptotic behavior for ferromagnetically and anti-ferromagnetically coupled modified Ising model of a network. For a ferromagnetically coupled \banet\, when $\tp \rightarrow \infty$, $\beta \cc \mg \kk_{i} \rightarrow 0$, so $\exp(-\beta \cc \mg \kk_{i}) \rightarrow 1 \implies \mg \rightarrow \frac{1}{2}$. As $\tp \rightarrow 0$, $\bt \cc \mg \kk_{i} \rightarrow \infty$, so $\exp(-\bt \cc \mg \kk_{i}) \rightarrow 0 \implies \mg \rightarrow 1$. These confirm the observations in the top panel in \fig\
\ref{fig:h0}. Similarly for an anti-ferromagnetically coupled \mim\ of a \banet\ we can verify the limit cases: as $\tp \rightarrow \infty, \bt \cc \mg \kk_{i} \rightarrow 0, \exp(\bt \cc \mg \kk_{i}) \rightarrow 1 \implies \mg \rightarrow \frac{1}{2}$. On the other hand, as $\tp \rightarrow 0$, $\exp(\bt \cc \mg \kk_{i}) \rightarrow \infty \implies \mg \rightarrow 0$. These validate the observations in the bottom panel of \fig\ \ref{fig:h0}. In order to compare the results of mean-field approximation with Monte Carlo simulations, we have plotted the results using these two different approaches in \fig\ \ref{fig:mft}.

For $\tp >> 1$, using Taylor expansion $\mg$ can be approximated as, $\mg \approx \frac{1}{2\pm\bt\cc\mm}$. We can conclude that, for a fixed large $\tp$ in ferromagnetically coupled systems, those with larger $\cc$ and $\mg$ have larger $\mg$ and vice versa. This investigation predicts the behavior of the system presented in \fig\ \ref{fig:comparison} and validates Monte Carlo simulations. The situation is reversed for an anti-ferromagnetically coupled system due to the presence of plus sign in the denominator. Eqs. \ref{eq:mfe1} and \ref{eq:mfe2} indicates that at $\tp >> 1$,

\begin{equation}
\mg \approx \frac{1}{2}\Bigg[ \frac{2 + \bt \hh}{2 \pm \bt \cc \mm}\Bigg]
\label{eq:mgtg1}
\end{equation}

However, in both cases, the asymptotic behavior of the system is preserved, for $\tp \rightarrow \infty$ (or $\bt \rightarrow 0), \mg \rightarrow \frac{1}{2}$ (\cf\ \eq\ \ref{eq:mgtg1}).  In the case where $\tp$ not tending to $\infty$, the value of $\mg$ depends on the magnitude and direction of magnetic field, $\hh$. This implies that for an anti-ferromagnetically coupled system, when $\hh > \cc \mm$ then $\mm > \frac{1}{2}$, however, for $\hh < \cc\mm$, $\mm > \frac{1}{2}$. Similar conclusions can be made when $\hh$ is negative in a ferromagnetically coupled system. Therefore, the behavior of the system changes at $|\bc| =\cc\mm$. 

\begin{figure}[!htb]
\textbf{(A)}\includegraphics[scale = 0.35]{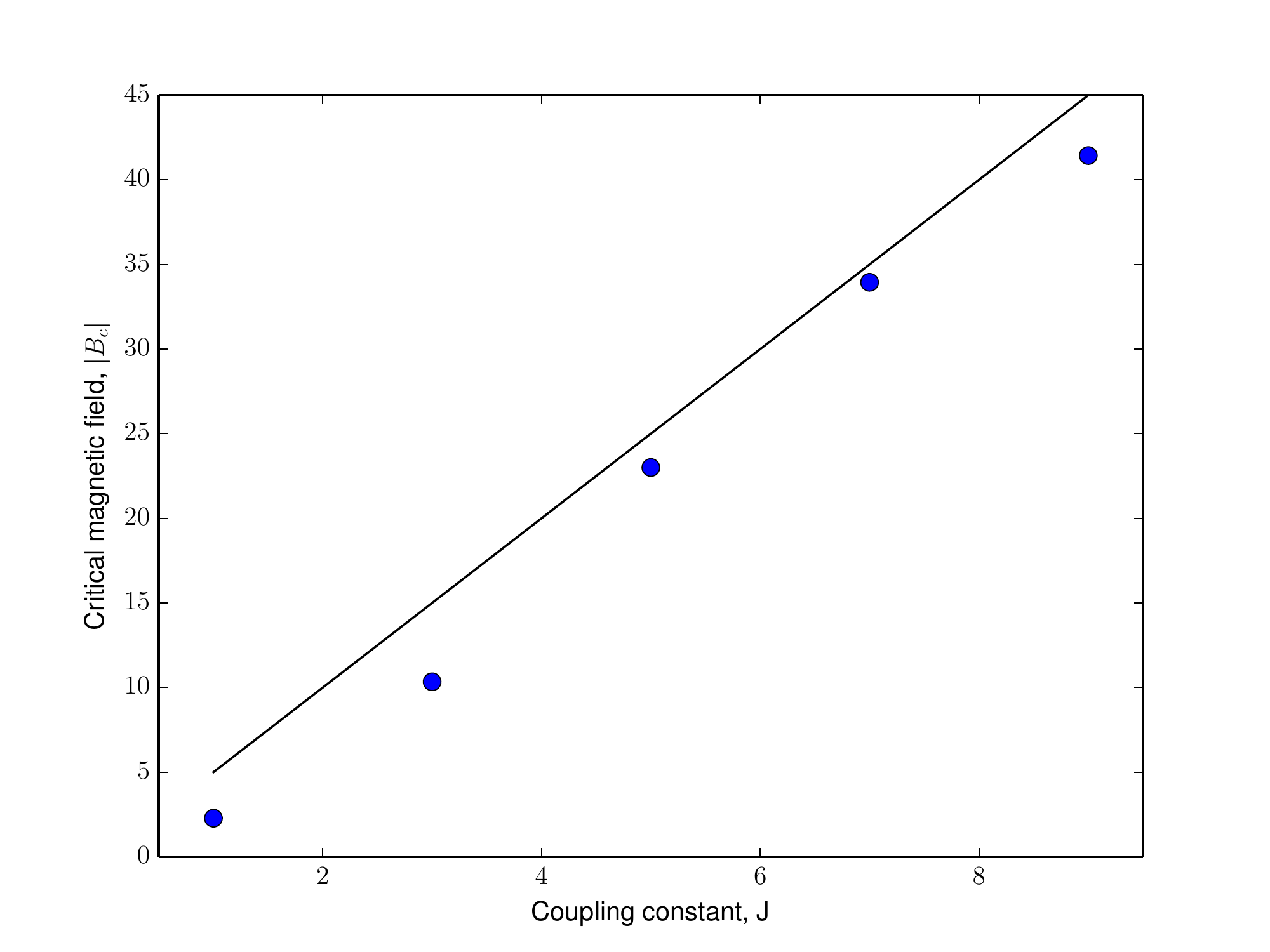}
\textbf{(B)}\includegraphics[scale = 0.35]{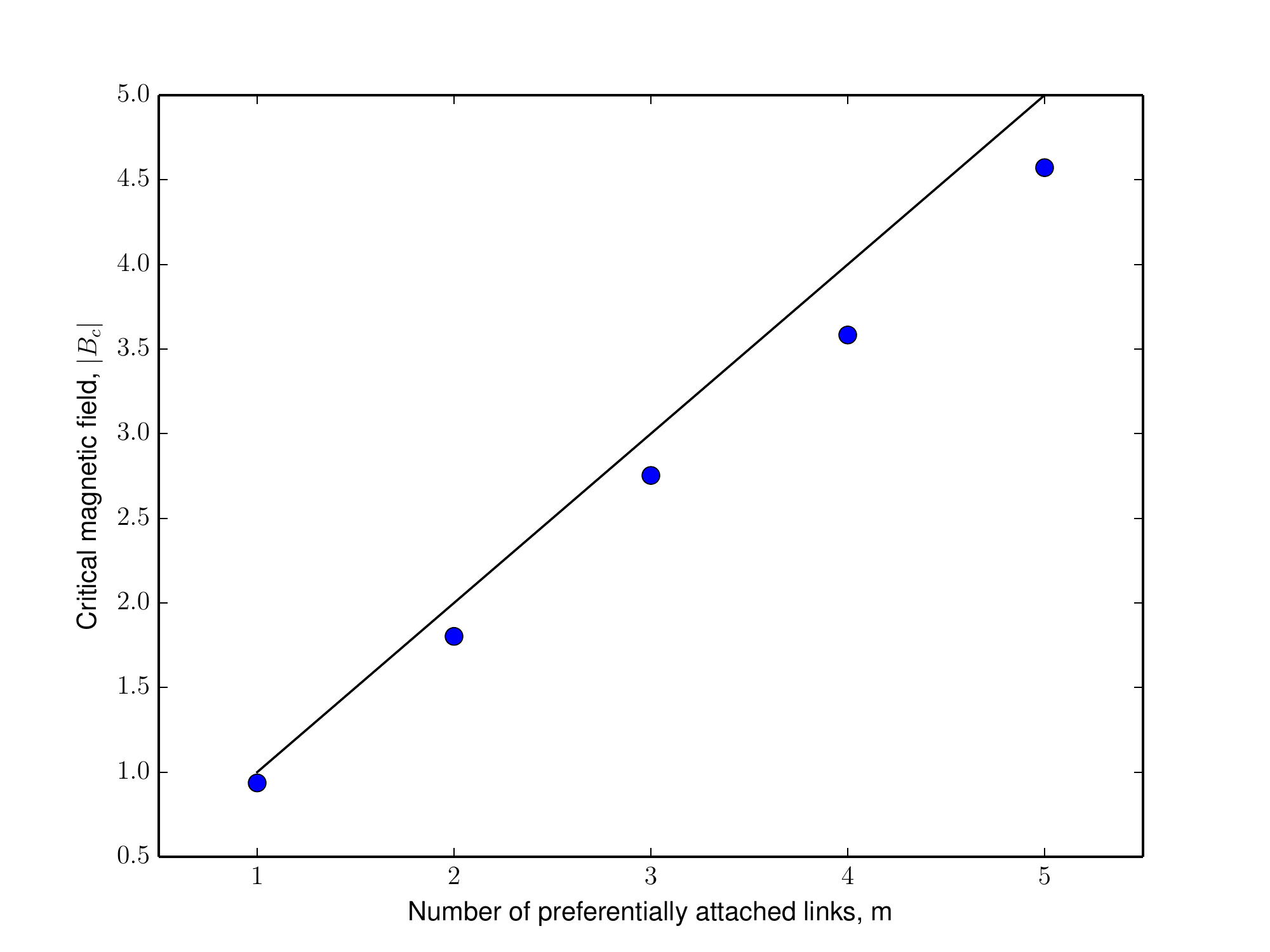}\\
\textbf{(C)}\includegraphics[scale = 0.35]{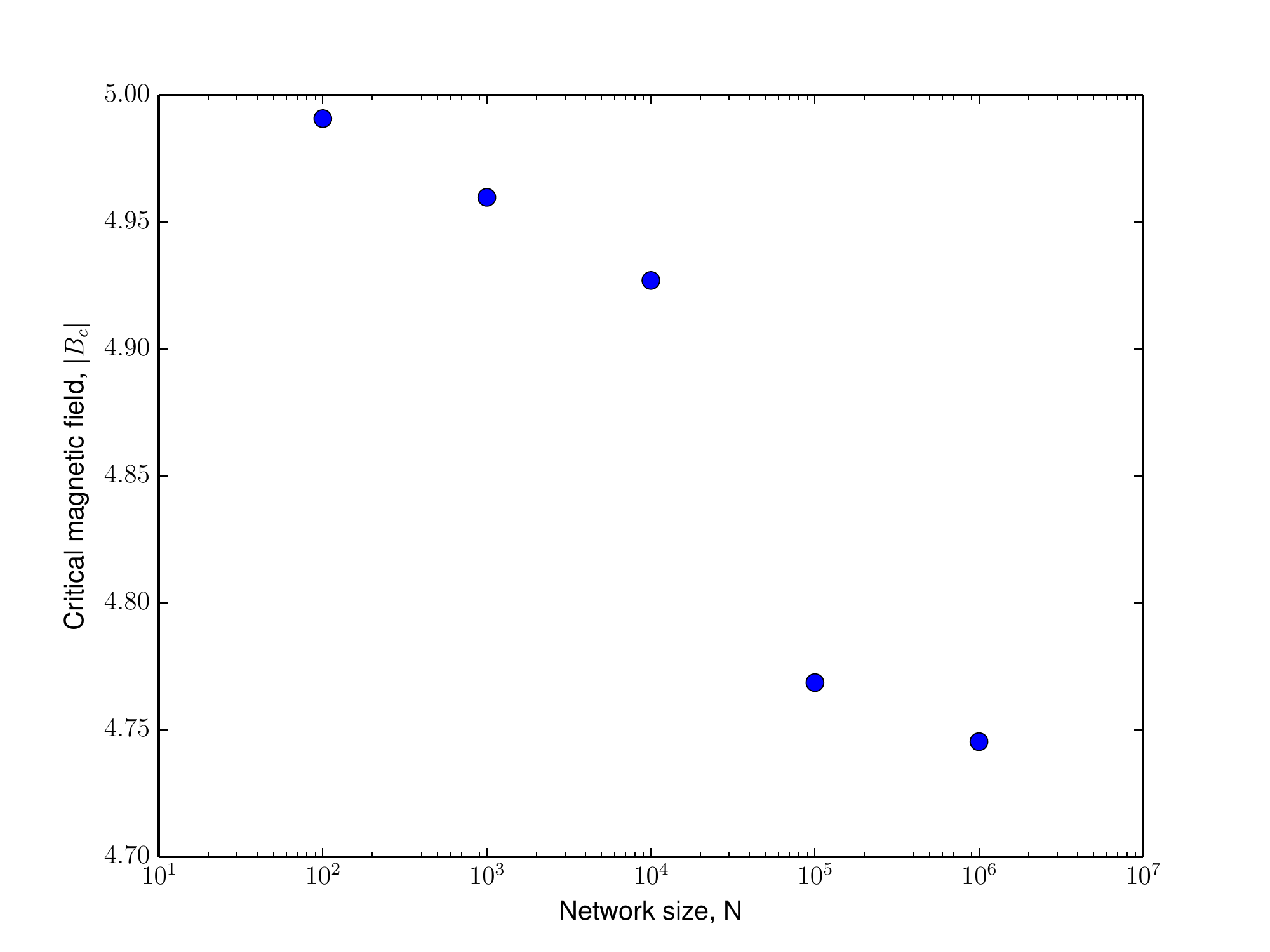}
\caption{Dependence of critical magnetic field, $\bc$ on network parameters for a \mim\ of a \banet\ of size $\nn = 5 \times 10^3$ with positive coupling constant, $\cc$. \textbf{(A)} Coupling constant, $\cc$ with fixed $\mm = 5$. \textbf{(B)} Number of preferentially attached links to construct \banet, $\mm$ with fixed $\cc = 1$. \textbf{(C)} on network size, $\nn$ with other simulation parameters fixed to $\cc = 1$ and $\mm = 5$. Blue dots indicate results from Monte Carlo simulations and black line indicates analytical results.}
\label{fig:scaling}
\end{figure}

This approximates the critical magnetic field, $\bc \approx 5$ for the choice of simulation parameters, which is very close to our observations from numerical simulations as can be verified in \figs\ \ref{fig:hg0} and \ref{fig:hl0}. Although the analytical results predict that the network size does not influence phase transition in the modified Ising model of the \banet, the numerical results predict a weak dependence of $\bc$ on network size (\fig\ \ref{fig:scaling}(C)), which appears in systems with large network sizes. The dependence on parameters $\cc$ and $\mm$ is over-estimated by the mean-field calculations as can be seen in \fig\ \ref{fig:comparison}. In the next \subsect\ \ref{subsec:mapping} we will derive the expression for the critical magnetic field by mapping the modified Ising spin system to the classical spin system on a \banet. 

\clearpage
\subsection[Mapping]{Mapping the \mim\ of \banet\ to classical Ising model of \banet}
\label{subsec:mapping}
The numerical and analytical observations presented in \sect\ \ref{sec:numerical} and \ref{sec:analytical_methods} can be validated by mapping the Hamiltonian of the \mim\ of \banet\, $\hm_{\gs}$ to the well-established classical Ising spin system on \banet\, $\hm_{\ms}$. Rewriting the modified Ising model \eq\ \ref{eq:hamiltonian},

\begin{equation}
\hm_{\gs} = - \frac{1}{2}\sum_{i,j}^{\nn}\cm\si\sj - \hh \sum_{i=1}^{\nn}\si \hspace{0.5cm} \si = \gs
\label{eq:rhm}
\end{equation}

\noindent This can be mapped to the Hamiltonian of the classical spin system by introducing new spin variables as,

\begin{equation}
\ns = 2 \Bigg(\si - \frac{1}{2}\Bigg)
\label{eq:mpspin}
\end{equation}

\noindent For $\si = 0 \rightarrow \ns = -1$ and for $\si = 1 \rightarrow \ns = 1$. Substituting the spin variables in the Hamiltonian \eq\ \ref{eq:rhm} we make the $\hm_{\gs} \rightarrow \hm_{\ms}$ transformation,
 
\begin{equation}
\begin{split}
\hm_{\gs} &= \frac{-1}{2}\sum_{i,j}^{\nn}\cm\si\sj-\hh\sum_{i=1}^{\nn}\si \hspace{0.5cm} \si =\gs\\
\hm_{\ms} &= \frac{-1}{2}\cm \Big( \frac{\ns + 1}{2} \Big) \Big( \frac{\nsj + 1}{2} \Big) - \hh \sum_{i=1}^{\nn} \Big( \frac{\ns + 1}{2} \Big) \hspace{0.5cm} s^{\prime} = \ms \\
&= \frac{-1}{2} \sum_{i,j}^{\nn} \frac{\cm}{4}\ns\nsj -
\frac{1}{2}\sum_{i,j}^{\nn} \frac{\cm}{4} (\ns + \nsj) -\frac{1}{2}\sum_{i,j}^{\nn}\frac{\cm}{4}-\frac{\hh}{2}\sum_{i=1}^{\nn}\ns - \frac{\hh}{2}\\
\label{eq:mapping1}
\end{split}
\end{equation}

\noindent Since $\cm = J_{ji}$, $\sum_{i,j}^{\nn}(\ns + \nsj) = 2\sum_{i,j}^{\nn}\ns$, eq. \ref{eq:mapping1} can be re-written as,

\begin{equation}
\begin{split}
\hm_{\ms} &= -\frac{1}{2}\sum_{i,j}^{\nn}\frac{\cm}{4}\ns\nsj - \sum_{i,j}^{\nn} \frac{\cm}{2}\si - \sum_{i,j}^{\nn}\frac{\cm}{8}-\frac{\hh}{2}\sum_{i=1}^{\nn}\ns - \frac{\hh}{2}\\
&= -\frac{1}{2}\sum_{i,j}^{\nn} \underbrace{\frac{\cm}{4}}_{\mathrm{new\ coupling, \cm^{\prime}}}\ns\nsj 
- \sum_{i=1}^{\nn} \underbrace{\Big[ \frac{\hh}{2} + \sum_{j=1}^{\nn} \frac{\cm}{2}\Big]}_{\mathrm{new\ local\ magnetic\ field,\ \nh}}\ns - \underbrace{\Big[ \sum_{i,j}^{\nn} \frac{\cm}{8} + \frac{\hh}{2}\Big]}_{\mathrm{constant,} E_0}\\
\label{eq:mapping2}
\end{split}
\end{equation}

\noindent So the problem of an Ising model with gene-type spin system is mapped on to a problem of Ising model with classical spin system as,

\begin{equation}
\hm_{\ms} = E_0 - \frac{1}{2} \sum_{i,j}^{\nn}\cm^{\prime}\ns\nsj -  \sum_{i=1}^{\nn}\nh_{i}\ns
\label{eq:mapH1}
\end{equation}

\noindent where constant $E_0 = -\sum_{i,j}^{\nn} \frac{\cm}{8}- \frac{\hh}{2}$, new coupling $\cm^{\prime} = \frac{\cm}{4}$ and new local magnetic field, $\nh = \frac{\hh}{2} + \sum_{j=1}^{\nn} \frac{\cm}{2}$. The fact that even in the absence of magnetic field there is an intrinsic local magnetic field, a $\sum_{i,j}^{\nn} \frac{\cm}{2}$ in the system reflects the asymmetricity of the spins present in the problem. In principle, any physical quantity of the system of modified Ising model of a interaction can therefore be derived from the system of Ising spins,

\begin{equation}
\z_{\gs}(\cm, \hh) = e^{\bt E_0}\z_{\ms}(\cm^{\prime}, \nh_{i})
\label{eq:zmap}
\end{equation}

\noindent However we are interested in the critical magnetic field as derived in \subsect\ \ref{subsec:mft}. Note that, the first term of the right hand of \eq\ \ref{eq:mapH1} is a constant and by redefinition of the zero of energy we have, 

\begin{equation}
\hm = -\frac{1}{2}\sum_{i,j}^{\nn}\cm^{\prime}\ns\nsj - \sum_{i=1}^{\nn}\nh_{i}\ns
\label{eq:mapH2}
\end{equation}

\noindent This is the Hamiltonian for a chosen realization of the network. So the ensemble average of the system Hamiltonian is,

\begin{equation}
\langle \hmf \rangle = -\frac{1}{2}\sum_{i,j}^{\nn}\langle \cm^{\prime} \rangle \ns\nsj - \sum_{i=1}^{\nn}\langle\nh_{i}\rangle \ns
\label{eq:hmfavg}
\end{equation}

\noindent where,

\begin{equation}
\begin{split}
\langle \nh_{i} \rangle &= \frac{\hh}{2} + \frac{-\cc}{2}\sum_{j=1}^{\nn}\adj\\
&= \frac{\hh}{2} - \frac{\cc}{2}\sum_{j=1}^{\nn}\frac{\kk_i \kk_j}{2\mm\nn}\\
&= \frac{\hh}{2}-\frac{\cc}{2}\kk_i\\
\label{eq:nh}
\end{split}
\end{equation}

\noindent The average critical field for the system $\bc$ can be derived by,

\begin{equation}
\frac{\bc}{2}-\frac{\cc}{2}\mk = 0
\label{eq:bc1}
\end{equation}

\noindent where $\kk_i$ is approximated by the average number of links, $\mk$. Note that $\mk = \frac{1}{\nn}\sum_{i=1}^{\nn}\kk_i = \frac{1}{\nn}\times 2\mm\nn = 2\mm$, thus,

\begin{equation}
\bc \approx \cc \mm
\label{eq:bc2}
\end{equation}

\noindent This validates our results presented in \sect\ \ref{subsec:mft}. \eq\ \ref{eq:bc2} predicts that the critical magnetic field depends linearly on $\cc$ and $\mm$. The numerical simulations confirms the analytical predictions on critical magnetic field (\fig\ \ref{fig:scaling}(A) and (B)). 
\section{Conclusions}
\label{sec:conclusions}

In living systems, it has been known that collective flipping of coherently expressed genes is associated with disease progression. This flipping causes the step by step change in the phenotype of the cell, causing it to transition from normal phase to diseased phase. Similarly, in magnetic systems, it has been known that collective flipping of spins is associated with the loss of spontaneous magnetization. Therefore it is intuitive to consider gene networks as two-state thermodynamic systems in a heat bath obeying Boltzmann statistics. 

In this regard, we have proposed here an adaptation of a well-established model in statistical mechanics that could be used to study phase transitions in living systems. This is a general statistical method to deal with non-linear large scale models arising in the context of biological networks and is scalable to any network size. We have presented a basic numerical and theoretical framework to investigate scale-free networks whose activity is modeled by a binary random variable. Taking the \bamod\ as the toy model, we have shown that the critical magnetic field, $\bc$ of the system scales linearly as a function of the number of preferentially attached links, $\mm$ and coupling constant $\cc$. Such a system undergoes a discontinuous phase transition of the first-order and exhibits hysteresis. Further, we have shown that the \mim\ can be mapped to a classical Ising model of a \banet. The simulation setup presented herein can be directly used for any biological network connectivity dataset and is also applicable to other networks that exhibit similar states of activity.

There are a few caveats to the analysis presented in the context of the \mim. We make the assumption that the gene-gene connectivity matrix is a binary matrix. This could instead be a matrix of varying degrees of connection strengths. Also, though it is known that gene expression is approximately a bimodal distribution and the Ising model two-state approximation is not far from reality, perhaps it is worth generalizing the \mim\ to a continuum of activity states such as in a Potts model.

\section{Appendix}
\label{sec:appdx}

Here we summarize the approach from \cite{bianconi_2002} to reduce mean adjacency matrix over many realization of \banet\ to network parameters. Let us consider a \banet\ of $\nn$ nodes. Starting from a small number of nodes $n_{0}$ and links $\mm_{0}$ (where $n_{0}, \mm_{0} << \nn$), the network is constructed iteratively by the constant addition of nodes with $\mm$ links. The new links are preferentially attached to well connected nodes in such a way that at time $t_j$, the probability $\pij$ that the new node $j$ is linked to node $i$ with connectivity $k_i(t_j)$ is given by,

\begin{equation}
\pij = \mm \frac{k_{i}(t_j)}{\sum_{\alpha = 1}^{j} \kk_{\alpha}}
\label{eq:connprob1}
\end{equation}

is proportional to the number of links $\kk_i$ at time $t_j$, and number of preferentially attached links $\mm$. The dynamic solution of connectivity at time $t_i$ is,

\begin{equation}
\kk_{i} = \mm \sqrt{\frac{t}{t_i}}
\label{eq:kt}
\end{equation}

From \eqs \ref{eq:connprob1} and \ref{eq:kt} we have,

\begin{equation}
\pij = \mm \frac{\mm \sqrt{\frac{t}{t_i}}}{\sum_{\alpha = 1}^{j} k_{\alpha}(t)}
\label{eq:connprob2}
\end{equation}

If $\nn$ is large we can approximate the total number of edges in the network at time $t_j$, given by the sum $\sum_{\alpha=1}^{j}\kk_{\alpha}$ as, 

\begin{equation}
\sum_{\alpha = 1}^{j} \kk_{\alpha} = \mm_0 + 2\mm t_j \approx 2\mm t_j
\label{eq:kalpha} 
\end{equation}

because $\mm_{0} << \nn$. The factor $2$ comes from the fact that as we create a link which connects two nodes, the number of links of each of them increases by $1$. Substituting \eq\ \ref{eq:kalpha} in \ref{eq:connprob2},

\begin{equation}
\begin{split}
\pij &= \frac{\mm^2 \sqrt{\frac{t_j}{t_1}}}{2\mm t_j}\\
&= \frac{\mm}{2} \frac{1}{\sqrt{t_it_j}}\\
\label{eq:connprob3}
\end{split}
\end{equation} 

The adjacency elements of the network $\adj$ are equal to $1$ if there is a link between node $i$ and $j$ and $0$ otherwise. Consequently the mean over many copies of a \banet\,

\begin{equation}
\langle \adj \rangle = \pij = \frac{\mm}{2}\frac{1}{\sqrt{t_it_j}}
\label{eq:adjmean}
\end{equation}

From \eq\ \ref{eq:kt} we can re-write for $t = \nn$ steps,

\begin{equation}
\begin{split}
\kk_{i}{(t)} &= \mm \sqrt{\frac{t}{t_i}}\\
\kk_{i}(\nn) &= \mm \sqrt{\frac{\nn}{t_i}}\\
t_i &= \frac{\mm^2 \nn}{\kk_{i}^{2}}
\label{eq:ti}
\end{split}
\end{equation}

and similarly,

\begin{equation}
t_j = \frac{\mm^2\nn}{\kk_{j}^{2}}
\label{eq:tj}
\end{equation}

From \eqs\ \ref{eq:ti} and \ref{eq:tj}, 

\begin{equation}
\begin{split}
\langle \adj \rangle &= \frac{\mm}{2} \frac{1}{\sqrt{\frac{\mm^2\nn}{\kk_{i}^{2}}}\sqrt{\frac{\mm^2\nn}{\kk_{j}^{2}}}}\\
&= \frac{1}{2\mm\nn}\kk_{i}\kk_{j}\\
\label{eq:adjnp1}
\end{split}
\end{equation}

The average of the adjacency matrix over many realizations can be approximated by the network parameters as,

\begin{equation}
\langle \adj \rangle = \frac{1}{2\mm\nn}\kk_{i}\kk_{j}
\label{eq:adjnp2}
\end{equation}

\section*{Funding}

\noindent This work was supported by the Deutsche Forschungsgemeinschaft (DFG) through GSC111; and Exploratory Research Space (ERS) Seed Fund 2017 in Computational Life Sciences (CLS001). All simulations were performed using the RWTH Compute Cluster under general use category; priority category allocated to AICES and JRC users; and with specific computing resources granted by RWTH Aachen University under project rwth0348. The authors gratefully acknowledge the generous support of the aforementioned funding and computing resources.


\begin{thebibliography}{99}

\bibitem[{Albert(2002)}]{albert_2002}
Albert,~R., Barabasi,~A.-L.
\newblock \emph{Statistical mechanics of complex networks}
\newblock Rev. Mod. Phys. 74, 47 (2002)

\bibitem[{Aleksiejuk(2002)}]{aleksiejuk_2002}
Aleksiejuk,~A., Holyst, ~J. ~A. , Stauffer, ~D.
\newblock \emph{Ferromagnetic phase transition in Barab'{a}si-Albert networks}
\newblock Physica A 310, 260–266 (2002)

\bibitem[{Bianconi(2002)}]{bianconi_2002} 
Bianconi, ~G.
\newblock \emph{Mean field solution of the Ising model on a Barab\'{a}si–Albert network}
\newblock Physics Letters A 303 (2002) 166–168

\bibitem[{Barrat(2000)}]{barrat_2000}
Barrat, A. and Weigt, M. 
\newblock \emph{On the properties of small-world network models}
\newblock Eur. Phys. J. B 13, 547 (2000)

\bibitem[{Castellano(2009)}]{castellano_2009}
Castellano, ~C., Fortunato, ~S., Loreto, ~V.
\newblock \emph{Statistical physics of social dynamics}
\newblock Reviews of Modern Physics. 2009; 81(2):591–646.

\bibitem[{Davies(2011)}]{davies_2011}
Davies,~P., Demetrius,~L., Tuszynski, ~J.~A.
\newblock \emph{Cancer as a Dynamical Phase Transition}
\newblock Theoretical Biology and Medical Modelling 2011, 8:30

\bibitem[{Dorogovstev(2002)}]{dorogovstev_2002}
Dorogovtsev, ~S. ~N., Godtsev, ~A. ~V. , Mendes, ~J. ~F. ~F.
\newblock \emph{Ising Model on Networks with an Arbitrary Distribution of Connections}
\newblock 	10.1103/PhysRevE.66.016104 (2002)

\bibitem[{Dorogovstev(2008)}]{dorogovstev_2007}
Dorogovtsev, ~S. ~N., Godtsev, ~A. ~V. , Mendes, ~J. ~F. ~F.
\newblock \emph{Critical phenomena in complex networks}
\newblock Rev. Mod. Phys. 80, 1275–1335 (2008)

\bibitem[{Facciotti(2013)}]{facciotti_2013}
Facciotti, ~M.~T.
\newblock \emph{Thermodynamically inspired classifier for molecular phenotypes of health and disease}
\newblock PNAS, vol.110:48 (2013)

\bibitem[{Ferreira(2010)}]{ferreira_2010}
Ferreira, ~A.~L., Mendes, ~J.~F.~F., Ostilli, ~M.
\newblock \emph{First- and second-order phase transitions in Ising models on small world networks, simulations and comparison with an effective field theory}
\newblock 	arXiv:1001.1342 (2010)

\bibitem[{Herrero(2008)}]{herrero_2008}
Herrero, ~.C.~P.
\newblock \emph{Anti-ferromagnetic Ising model in small-world networks}
\newblock Phys. Rev. E, 77, 041102, (2008)

\bibitem[{Herrero(2002)}]{herrero_2002}
Herrero, ~C.~P.
\newblock \emph{Ising model in small-world networks}
\newblock Phys. Rev. E 65, 066110 (2002)

\bibitem[{Ising(1925)}]{ising_1925}
Ising, ~E. (1925). 
\newblock \emph{Beitrag zur Theorie des Ferromagnetisms}.
\newblock Z.~Phys~, pp.\ v.~31, 253

\bibitem[{Gitterman(2000)}]{gitterman_2000}
Gitterman, ~M.
\newblock \emph{Small-world phenomena in physics: the Ising model}
\newblock J. Phys. A 33, 8373 (2000)

\bibitem[{Lopes(2004)}]{lopes_2004}
Lopes, J. V. , Pogorelov, Y. G., dos Santos, J. M. B. L.
\newblock \emph{Exact Solution of Ising Model on a Small-World Network}
\newblock cond-mat/0402138 (2004)

\bibitem[(Torquato(2010))]{torquato_2010}
S. Torquato
\newblock \emph{Toward an Ising Model of Cancer and Beyond}
\newblock \emph{arXiv:1010.6284v2 [q-bio.CB]}

\bibitem[{Metropolis(1953)}]{metropolis_1953}
Metropolis, N., Rosenbluth, A. ~W, Rosenbluth, M. ~N, Teller, A. ~H and Teller, E. (1953)
\newblock \emph{Equation of State Calculations by Fast Computing Machines}
\newblock \emph{J. Chem. Phys.}, 21~, 1087

\bibitem[{Mojtahedi(2016)}]{mojtahedi2016}
Mojthahedi, ~M., Skupin, ~A., Zhou, ~J., Castano, ~I.~G., Leong-Quong, ~.Y.~R.,
Chang, ~H., Trachana, ~K., Giuliani, ~A., Huang, ~S.
\newblock \emph{Cell Fate Decision as High-Dimensional Critical State Transition
}
\newblock PLoS Biol 14(12): e2000640.

\bibitem[{Pastor(2015)}]{pastor_2015}
Pastor-Satorras, ~R., Castellano, ~C., Mieghem, ~P. ~V. , Vespignani, ~A.
\newblock \emph{Epidemic processes in complex networks}
\newblock Rev. Mod. Phys. 87, 925–979 (2015)

\bibitem[{Pekalski(2001)}]{pekalski_2001}
Pekalski, ~A.
\newblock \emph{Ising model on a small-world network}
\newblock Phys. Rev. E 64, 057104 (2001)

\bibitem[{Scheffer(2001)}]{scheffer_2001}
Scheffer, ~M., Carpenter, ~S., Foley, ~J.~A., Folke,~C., Walker, ~B.
\newblock \emph{Catastrophic shifts in ecosystems}
\newblock Nature 2001, 413:591-596.

\bibitem[{Scheffer(2012)}]{scheffer_2012}
Scheffer, ~M., Carpenter, ~S., Timothy, ~L., Bascompte, ~J., Brock, ~W., 
Dakos, ~V., van de Koppel, ~J., van de Leemput ~I.~A., Levin, ~S.~A., van Nes, ~E.,
Pascual, ~M., Vandermeer, ~J.
\newblock \emph{Anticipating Critical Transitions}
\newblock Science, Vol. 338 (2012)

\bibitem[{Aldana(2004)}]{aldana_2004}
Aldana, ~M., Larralde, ~H.
\newblock \emph{Phase transitions in scale-free neural networks: Departure from the standard mean-field
universality class
}
\newblock Physical Review E 70, 066130 (2004)

\bibitem[{Stauffer(2008)}]{stauffer_2008}		
D. Stauffer  (2008).
\newblock \emph{Social applications of two-dimensional Ising models}
\newblock Am. J. Phys. 76 (2008) 470.

\bibitem[{Smith(2010)}]{smith_2010}
Smith, ~A.~S.   
\newblock \emph{Physics Challenged by Cells}
\newblock Nature Physics, 6:726-729 (2010)

\bibitem[{Onsager(1944)}]{Onsager1944}
Onsager, ~L.
\newblock \emph{Crystal statistics. I. A two-dimensional model with an order-disorder transition}
\newblock Physical Review, Series II 65(3–4):117–149 (1944)

\bibitem[{May(2001)}]{may_2001}
May, ~R. ~M., Lloyd, ~A. ~L.
\newblock \emph{Infection dynamics on scale-free networks}
\newblock Physical Review Letters E, 64 (2001)

\bibitem[{Pastor(2001)}]{pastor_2001}
Pastor-Satorras, ~R., Vespignani, ~A.
\newblock \emph{Epidemic Spreading in Scale-Free Networks}
\newblock Physical Review Letters E (2001)

\bibitem[{Bartolozzi(2006)}]{bartolozzi_2006}
Bartolozzi, ~M. and Surungan, ~T. and Leinweber, ~D. ~B. and Williams, ~A. ~G.
\newblock \emph{Spin-glass behavior of the antiferromagnetic Ising model on a scale-free network}
\newblock Physical Review B - Condensed Matter and Materials Physics 73:1--19 (2006)

\bibitem[{Contucci(2007)}]{contucci_2007}
Contucci, ~P.,Ghirlanda, ~S,
\newblock \emph{Modeling society with statistical mechanics: an application to cultural contact and immigration}
\newblock Qual. Quantit., 41:569 --578 (2007)

\bibitem[{Kumar(2000)}]{kumar_2000},
Kumar, ~R., Raghavan, ~P., Rajagopalan, ~D., Sivakumar, ~D., Tomkins, ~A., Upfal, E
\newblock \emph{The Web as a graph}
\newblock Proceeding of the 9th ACM Symposium on Principles of Database Systems (2000)

\bibitem[{Stauffer(2006)}]{stauffer_2006},
Stauffer, ~D., Hohnisch, ~M., Pittnauer, S
\newblock \emph{The impact of external events on the emergence of social herding of economic sentiment}
\newblock Physica A 370 (2006)

\bibitem[{Holstein(2013)}]{holstein_2013}
Holstein, ~D., Goltsv, ~A. ~V., Mendes, ~J. ~F. ~F.
\newblock \emph{Impact of noise and damage on collective dynamics of scale-free neuronal networks}
\newblock Phys. Rev. E 87 (2013)

\bibitem[{Pastor(2015)}]{pastor_2015}
Pastor-Satorras, ~R., Castellano, ~C., Van Mieghem, ~P., Vespignani, ~A.
\newblock \emph{Epidemic Processes in Complex Networks}
\newblock arXiv:1408.2701v2 (2015)

\bibitem[{Scheffer(2012)}]{scheffer_2012}
Scheffer, ~M. and Carpenter, ~S. ~R., Lenton, ~T. ~M., Bascompte, ~J., Brock, ~W., Dakos, ~V., van de Koppel, ~J. and van de Leemput, ~I. A., Levin, ~S. ~A., van Nes, ~E. ~H., Pascual, ~M., Vandermeer, ~J.
\newblock \emph{Anticipating Critical Transitions}
\newblock Science 6105: 338: 344--348  (2012)

\bibitem[{Razquin(2018)}]{razquin_2018}
Cesar-Razquin, ~A., Girardi, ~E.,Yang, ~M.,Brehme, M., Saez-Rodriguez, ~J., Superti-Furga, ~G.
\newblock \emph{In silico Prioritization of Transporter–Drug Relationships From Drug Sensitivity Screens}
\newblock Front. Pharmacol., doi: 10.3389/fphar.2018.01011 (2018)

\bibitem[{Jin(2017)}]{jin_2017}
Jin, ~B., Liu, ~R., Hao, ~S., Li, ~Z., Zhu, ~C., Zhou, ~X.
\newblock \emph{Defining and characterizing the critical transition state prior to the type 2 diabetes disease}
\newblock PLoS One 12 (2017)

\bibitem[{Liu(2013)}]{liu_2013}
Liu, ~X., Liu, ~R., Zhao, ~X. ~M., Chen, ~L.
\newblock \emph{Detecting early-warning signals of type I diabetes and its leading biomolecular networks by dynamical network biomarkers}
\newblock BMC Medical Genomics 6 (2013)

\bibitem[{Mojtahedi(2016)}]{mojtahedi_2016}
Mojtahedi, ~M., Skupin, ~A., Zhou, ~J., Castano, ~I. ~G., Leong-Quong, ~Y. ~R., Chang, ~H., Trachana, ~K., Giuliani, ~A., Huang, ~S.
\newblock \emph{Cell Fate Decision as High-Dimensional Critical State Transition}
\newblock PLoS Biology 14 (2016)

\bibitem[{Smith(2010)}]{smith_2010}
Smith, ~A. ~S.
\newblock \emph{Physics Challenged by Cells}
\newblock Nature Physics 6:726-729

\bibitem[{Trefois(2015)}]{trefois_2015}
Trefois, ~C., Antony, ~P.~M.~A., Goncalves, ~J., Skupin, ~A., Balling, ~R.
\newblock \emph{Critical transitions in chronic disease: Transferring concepts from ecology to systems medicine}
\newblock Current Opinion in Biotechnology 34:48--55 (2015)

\bibitem[{Torquato(2010)}]{torquato_2010}
Torquato, ~S.
\newblock \emph{Towards an Ising Model of Cancer and Beyond}
\newblock arxiv:1010.6284v2

\bibitem[{Krishnan(2019)}]{krishnan_2019}
Krishnan, ~J., Torabi, ~R., Di Napoli, ~E., Schuppert, ~A.
\newblock \emph{A Statistical Mechanics Perspective of Phase Transitions in Living Systems}
\newblock SIAM Conference on Computational Science and Engineering (CSE) 2019, Spokane, Washington, USA

\bibitem[{Krishnan(2018)}]{krishnan_2018}
Krishnan, ~J., Torabi, ~R., Di Napoli, ~E., Schuppert, ~A.
\newblock \emph{Simulations of Phase Transitions in Living Systems}
\newblock Systems Biology of Human Diseases 2018, LA, USA 

\end{thebibliography}
\end{document}